\documentclass{aa}
\usepackage{graphicx}
\usepackage{txfonts}
\usepackage[colorlinks,citecolor=blue]{hyperref}
\usepackage{makecell}
\hypersetup{
    colorlinks = true,
    linkcolor = blue,
    anchorcolor = blue,
    citecolor = blue,
    filecolor = blue,
    urlcolor = blue
    }
\newcommand{\Teff}{\mbox{$T_{\mathrm{eff}}$}}
\begin{document} 
\newcommand{\red}[1]{\color{red}{#1}\color{black}}
\title{Faint southern spectrophotometric standard stars\thanks{Based on observations collected at the European Southern Observatory under ESO programmes 109.234T.001, 110.23NA.001, 111.255J.001 and 112.25PC.001.}}

   \author{N. P. Gentile Fusillo\inst{1,2}
          \and
           S. Moehler\inst{3}
          \and
           N. Przybilla\inst{4}
            \and 
           A. K. Elms\inst{5}
           \and
            P.-E. Tremblay\inst{5}      
          \and
          W. Kausch\inst{4}
          \and
          %E. Valenti\inst{3}
          %\and
          F. Kerber\inst{3}
          }

   \institute{Dipartimento di Fisica, Universita` di Trieste, Via Alfonso Valerio, 2, 34127 Trieste, Italy
   \and INAF-Osservatorio Astronomico di Trieste, Via G.B. Tiepolo 11, I-34143 Trieste, Italy\\
   \email{nicolapietro.gentilefusillo@units.it}
   \and
   European Southern Observatory, Karl-Schwarzschild-Stra\ss e, 2 85748 Garching bei M\"unchen, Germany
   \and
   Universit\"at Innsbruck, Institut f\"ur Astro- und Teilchenphysik, Technikerstr. 25/8, 6020 Innsbruck, Austria
    \and
    Department of Physics, University of Warwick, Coventry, CV4 7AL, UK}
   \date{Received ; accepted }

 \abstract
  {The advent of the Extremely Large Telescope (ELT) will increase the collecting area by more than an order of magnitude compared to the individual Unit Telescopes of the Very Large Telescope (VLT).
  Fainter spectrophotometric standard stars than those currently available in the $V$\,=\,11 to 13\,mag ($K$\,=\,12 to 14\,mag) range are required for spectroscopic observations with instruments such as the Multi-AO Imaging Camera for Deep Observations (MICADO) on the ELT, notably in the near-infrared~wavelength~regime.
 } 
   {We identify suitable spectrophotometric standard stars among white dwarfs with hydrogen atmospheres (DA white dwarfs) in the magnitude range $K$\,=14 to 16\,mag and provide reference data based on stellar model atmospheres. }
   {We observed 24 candidate DA white dwarfs with the X-shooter instrument on the VLT, covering the wavelength range 300\,nm to 2480\,nm in three arms. We took care to include stars at latitudes below and above $-25^\circ$ to allow observations for all wind directions at the location of the ELT. The spectra were analysed using model fluxes from 3D pure-hydrogen local thermodynamic equilibrium model atmospheres and multi-band photometry. From the sample of observed targets, we selected 14 reliable flux calibrators. For these targets, the residuals from the match between the model best-fit models and the observed spectra across the full wavelength range are $<$3\%, with the exception of the UV regions affected by the ozone Huggins bands (300\,nm--340\,nm) and regions contaminated by telluric lines.}
   {We have identified and fully characterised 14 DA white dwarfs that can be used as spectrophotometric standard stars for the MICADO instrument as well as any other future instrument with similar requirements in the brightness range, $K$\,=\,14 to 16\,mag (Vegamag), and provide reference fluxes.}

   {}
   \keywords{Standards -- Techniques: spectroscopic -- white dwarfs
               }

   \maketitle
%
%-------------------------------------------------------------------
\nolinenumbers
\section{Introduction}

Accurate spectroscopic flux calibrations of astronomical spectra can be used to reconstruct the spectral energy
distribution (SED) of astronomical objects, 
by removing instrumental signatures 
and recovering their absolute fluxes in physical units, both from within and outside
the Earth's atmosphere. This challenging process requires spectrophotometric standard stars (hereafter flux standard stars) with known absolute fluxes that are accessible to the same spectrograph with which the intended science spectra will be observed.

The development of telescopes of the $8-10$\,m class and the
drive towards higher spectral resolution and wider wavelength coverage (from the optical-UV to the IR) have reduced the usefulness of classical flux standards \citep[e.g.][]{Hamuy+92,Hamuy+94} because they do not extend to the near-infrared (NIR) and/or are too coarsely sampled to permit the flux calibration of high-resolution spectra. For instance, the extended wavelength coverage of the X-shooter instrument on the Very Large Telescope (VLT) of the European Southern Observatory (ESO) required an extension of flux standards towards the NIR \citep{Moehleretal14}. 

The much larger collective area of ESO's 39\,m Extremely Large Telescope \citep[ELT;][]{PaCi23} will require even fainter flux standard stars. Although recent work has significantly expanded the network of flux standards to include fainter objects, these still lack adequate coverage and/or sampling in the IR range, often providing model fluxes calibrated only on photometric SEDs \citep{Narayan2019,Axelrod2023,Bohlinetal2025}.   

This is particularly relevant for the spectroscopic mode of the Multi-AO Imaging Camera for Deep Observations \citep[MICADO;][]{Daviesetal16,Daviesetal18,Sturmetal24}, the first-light instrument of the ELT. Although MICADO is primarily designed for high-spatial-resolution imaging at high sensitivity across a field of view 
of up to almost a full arcminute-squared and high-contrast imaging, it will also provide a compact cross-dispersing slit spectrometer. It is designed to observe point and compact sources with a resolving power of up to $R$\,=$\lambda/\Delta\lambda$\,$\approx$\,20\,000, covering a wavelength range from the $I$ to $K$ bands in two setups, and providing slits of various widths and lengths. The sensitivity will be comparable to that of the \textit{James Webb} Space Telescope (JWST), but at six to seven times higher linear spatial and spectral resolution.

The spectroscopic mode of MICADO will suffer from significant slit losses, induced by its extremely narrow slits and the
broad wings of the point spread function (PSF), which also strongly depend on
atmospheric conditions that affect the quality of the adaptive optics (AO) correction. Slit losses in the $K$ band will be  about 50\%, rising
to some 80\% in the $Iz$ band under median atmospheric conditions, and to over 90\% under less favourable conditions.
This is for the best case, assuming a single conjugate adaptive optics (SCAO) PSF with the target used as a natural guide star (NGS) on-axis. Slit losses will increase for offsets between the target and the NGS and when using multi-conjugate adaptive optics (MCAO). Hot
flux standard stars with a steep SED rise  towards shorter wavelengths will counteract the effect, providing sufficient flux across the entire wavelength range covered. Hot white dwarfs with pure-hydrogen atmospheres (DA-type white dwarfs), which have classically been used as flux standard
stars, are perfectly suited for the task of flux-calibrating MICADO spectra, requiring similar
exposure times throughout the $I$ to $K$ bands. However, standard stars at $I$ to $K$ magnitudes between 13 and 16 with NIR coverage, as required for MICADO,
are essentially unavailable at the moment.

In this work we present a new set of calibrated model spectra for flux standard stars covering the wavelength range 340\,nm to 2480\,nm. These spectra are useful for deriving consistent instrumental response curves over this wide wavelength range with a spectral resolving power of up to 40\,000.

The paper is structured as follows. The target selection is discussed in Sect.~\ref{sec:sample}, and Sect.~\ref{sec:obs} concentrates on the observations and the data reduction. The modelling of the observed spectra via model atmospheres is described in Sect.~\ref{ssec:model_spectra}, and the calibration of the final set of flux standard stars is summarised in Sect.~\ref{sec:calib}. Finally, in Sect.~\ref{sec:conclusions}, conclusions are drawn and a strategy for future improvements is presented. 

\section{Target selection}\label{sec:sample}
The flux standard star candidates were selected from the 
catalogue of white dwarfs of \citet{GentileFusilloetal21} that was based on the \textit{Gaia} Early Data Release~3 \citep[EDR3;][]{Gaia16,GaiaEDR3} using the following criteria:
\begin{itemize}
    
\item A high probability of being a white dwarf ({\tt Pwd\,$>$\,0.75}), to select high-confidence white dwarf candidates.

\item A declination (Dec.) of $<$ $-$10$^\circ$, to have targets both north and south of  the ELT site such that suitable standards should be available even if pointing restrictions due to wind are in place.
\item A roughly uniform distribution across the full right ascension (RA) range, to ensure a suitable flux standard would be observable any time during the year.

\item An effective temperature range of 20\,000\,K\,$<$\,$T_\mathrm{eff}$\,$<$\,40\,000\,K (using the temperature estimates in \citealt{GentileFusilloetal21}), to have sufficient flux across the full wavelength range of {MICADO} even with its lower throughput at short wavelengths. Two stars below this $T_\mathrm{eff}$ limit were added at a later stage in order to complete the RA coverage. 
Though the most internally consistent set of flux standards, the \textit{Hubble} Space Telescope (HST) flux scale, relies only on hot DA white dwarfs \citep[$T_\mathrm{eff}$\,$>$\,30\,000\,K;][]{BohlinGordonTremblay2014}, recent work has shown the viability of DAs as cool as $\simeq 10\,000$\,K as flux standards \citep{gentilefusilloetal20,elmsetal24}. All targets were selected in a range of \Teff and $log$ g that excludes the nominal instability strip where DA white dwarfs undergo pulsations (ZZ Ceti; \citealt{vangrooteletal12}).

\begin{figure}[ht]
\includegraphics[angle=0, width=.99\columnwidth]{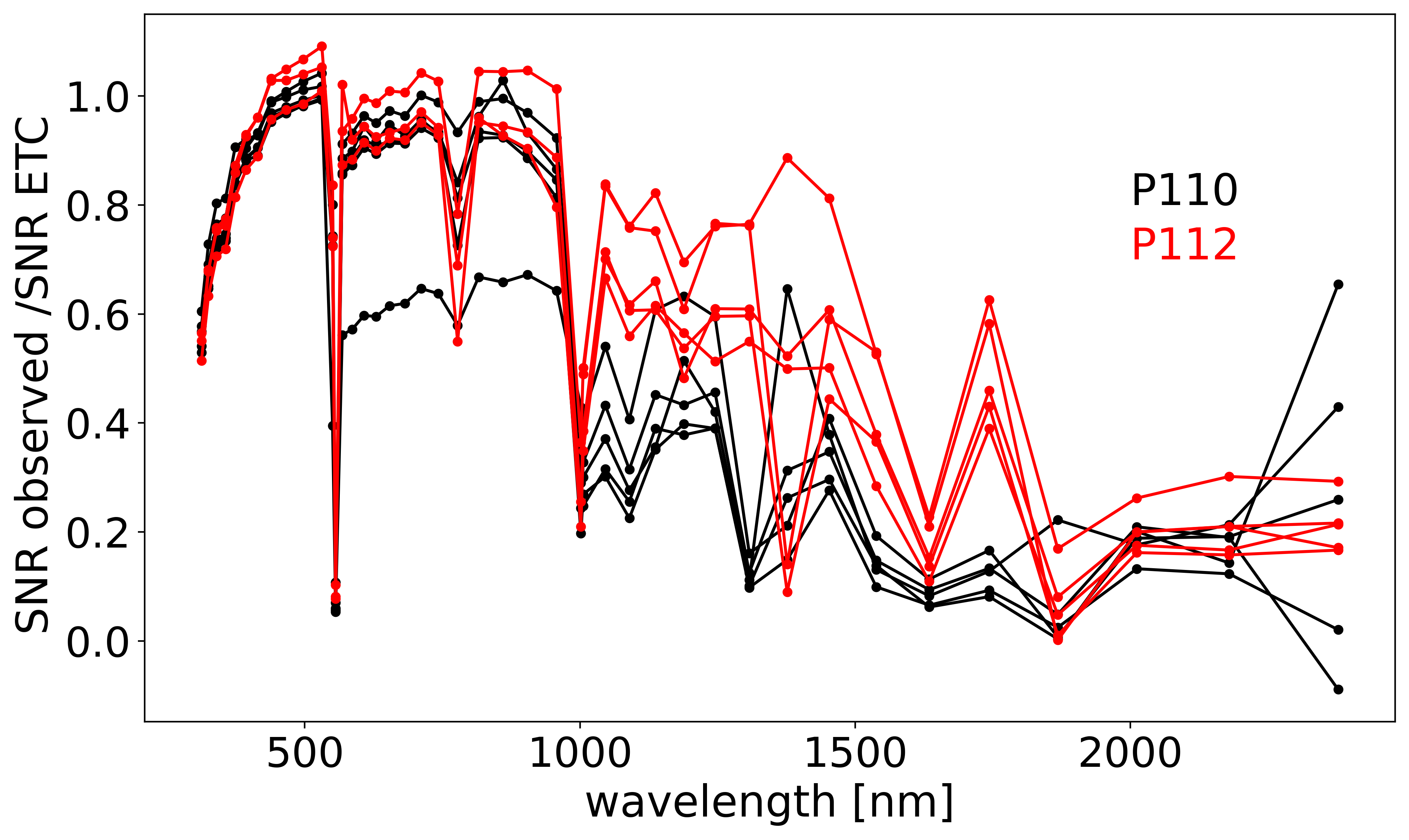}\\
\caption[]{Ratio of the S/N derived from the observed data by the X-shooter pipeline and the S/N predicted by the ETC for the actual observing conditions, at both blaze wavelengths of each order. The black lines show the results for the NIR data from P110 (slit width 5\arcsec), while the red lines show the results for P112 (slit width 1.2\arcsec). The blaze wavelengths 1376\,nm and 1867\,nm are strongly affected by telluric absorption, and the region around 1634\,nm has strong sky emission lines.}

\label{fig:NIR_SNR_ratio}
\end{figure}

\item DA candidates, to select the targets that can be most accurately modelled. This was achieved by first selecting objects located on the so-called A-branch of the white dwarf sequence \citep{GaiaColCMD18}. The selection was further refined by only including white dwarfs for which the pure-hydrogen model fit provided a smaller $\chi^2$ compared to the fit of other atmospheric models in \citet{GentileFusilloetal21}.

\item Isolated stars, to minimise the risk of flux contamination and exclude objects that would require a more difficult targeting procedure. This was achieved by excluding all objects with a neighbouring \textit{Gaia} source brighter than $G$\,=\,19\,mag within 10{\arcsec}. The selection was further refined by visually inspecting available images from multiple surveys. 

\item Non-variable stars. Only targets with the parameter {\tt excess\_flux\_error\,=\,0} in \citet{GentileFusilloetal21} were included, as higher values are potentially indicative of photometric variability.

\item Well-behaved SEDs. For each target, the SED was constructed using available photometry from the Panoramic Survey Telescope and Rapid Response System (Pan-STARRS,  \citealt{Chambersetal16}), SkyMapper \citep{Onkenetal24}, VLT Survey Telescope ATLAS (VLT ATLAS, \citealt{Shanksetal15}, Galaxy Evolution Explorer (GALEX,  \citealt{Bianchietal17}), Two Micron All Sky Survey (2MASS \citealt{Cutrietal03}), VISTA Hemisphere Survey (VHS, \citealt{McMahonetal21}), and (Wide-field Infrared Survey Explorer (WISE, \citealt{Cutrietal12}), and it was compared to the appropriate white dwarf model. Systems with SEDs significantly different from that of a single, isolated white dwarf were discarded as potentially hiding unresolved companions, circumstellar disks, strong magnetic fields, or, more generally, photometric variability.

\item Sufficient $K$-band brightness ($<$16\,mag). Whenever VHS or 2MASS $K$-band magnitudes were not available, synthetic $K$ magnitudes were computed from DA models corresponding to \Teff\ and surface gravities ($\log g$) from \citet{GentileFusilloetal21} appropriately scaled using the \textit{Gaia} DR3 parallax. \end{itemize}
The application of these selection criteria resulted in a sample of 24 southern flux standard candidates. 
As a final test, we checked the literature to ensure that whenever available,  previous classifications of these stars based on optical \citep{mccook&sion99, gianninasetal11, SPY2020} or UV \citep{sahuetal23} spectroscopy identified them as typical DA white dwarfs.

 \begin{table*}[ht]
\centering
\caption[]{Observational information on the initial set of 24 flux standard star candidates.}\label{tab:obs_info}
\setlength{\tabcolsep}{1.7mm}
\begin{tabular}{lllccrrr}
\hline
\hline
Target       & Run          & Start of           & \multicolumn{2}{c}{Range in}          & \multicolumn{3}{c}{Total exposure time} \\
             &              & observation         &  airmass           & seeing at zenith & UVB & VIS & NIR \\
             &              & [UT]                &                    & [arcsec]         & [sec] & [sec] & [sec]\\
             \hline

WDJ005317.43$-$325956.52 & 109.234T.001 & 2022-06-02T09:19:52 &  1.423\ldots 1.315 &  0.95\ldots 0.64 & 1080 & 1260 & 1500\\ 
WDJ010820.77$-$353441.70 & 110.23NA.001 & 2022-11-18T02:56:41 &  1.173\ldots 1.045 &  1.70\ldots 1.08 & 4500 & 4920 & 5400\\ 
WDJ013158.29$-$683114.51 & 110.23NA.001 & 2022-10-18T02:52:23 &  1.432\ldots 1.384 &  2.17\ldots 1.25 & 4500 & 4920 & 5400\\ 
WDJ014251.02$-$385906.76 & 112.25PC.001 & 2023-10-10T03:27:15 &  1.105\ldots 1.034 &  2.36\ldots 1.31 & 4560 & 4920 & 5400\\ 
WDJ022827.20$-$324233.80 & 112.25PC.001 & 2023-10-10T05:12:34 &  1.022\ldots 1.010 &  2.20\ldots 1.09 & 3360 & 3720 & 4200\\ 
WDJ042926.24$-$044845.62 & 112.25PC.001 & 2023-12-15T04:05:01 &  1.162\ldots 1.074 &  0.74\ldots 0.54 & 3360 & 3720 & 4200\\ 
WDJ051002.85$-$003755.65 & 112.25PC.001 & 2023-11-17T04:49:58 &  1.150\ldots 1.095 &  0.79\ldots 0.56 & 4560 & 4920 & 5400\\ 
WDJ060020.89$-$101404.50 & 110.23NA.001 & 2022-11-17T06:56:09 &  1.098\ldots 1.033 &  0.77\ldots 0.45 & 4500 & 4920 & 5400\\ 
WDJ064856.09$-$252346.96 & 110.23NA.001 & 2022-12-29T06:38:49 &  1.231\ldots 1.093 &  1.66\ldots 1.08 & 2760 & 3120 & 3600\\ 
WDJ090217.30$-$040655.48 & 112.25PC.001 & 2023-12-15T07:03:57 &  1.102\ldots 1.074 &  1.08\ldots 0.64 & 1840 & 2080 & 2400\\ 
WDJ101952.36$-$140734.26 & 110.23NA.001 & 2023-01-15T06:46:05 &  1.034\ldots 1.017 &  1.75\ldots 1.01 & 4560 & 4920 & 5400\\ 
WDJ103342.76$-$114138.34 & 110.23NA.001 & 2022-12-29T07:49:36 &  1.048\ldots 1.043 &  1.33\ldots 1.32 &  320 &  440 &  600\\ 
             &              & 2023-01-16T07:49:31 &  1.049\ldots 1.030 &  1.65\ldots 0.98 &  960 & 1320 & 1800\\ 
WDJ110112.24$-$131442.23 & 111.255J.001 & 2023-05-14T23:49:24 &  1.054\ldots 1.020 &  1.11\ldots 0.68 & 4500 & 4920 & 5400\\ 
WDJ131341.58$-$305133.71 & 111.255J.001 & 2023-04-14T05:33:25 &  1.205\ldots 1.046 &  0.90\ldots 0.56 & 4500 & 4920 & 5400\\ 
WDJ161419.13$-$083326.60 & 111.255J.001 & 2023-06-12T01:14:39 &  1.233\ldots 1.068 &  1.72\ldots 1.27 & 4500 & 4920 & 5400\\ 
WDJ161755.24$-$153552.15 & 109.234T.001 & 2022-04-06T08:10:43 &  1.021\ldots 1.015 &  2.03\ldots 1.55 &  500 &  720 &  700\\ 
             &              & 2022-05-11T07:35:14 &  1.202\ldots 1.138 &  1.18\ldots 0.89 & 1080 & 1260 & 1500\\ 
WDJ170625.36$-$762304.95 & 109.234T.001 & 2022-04-04T07:36:01 &  1.645\ldots 1.625 &  0.97\ldots 0.85 & 1720 & 1960 & 1000\\ 
             & 111.255J.001 & 2023-04-14T04:25:47 &  1.924\ldots 1.805 &  0.93\ldots 0.69 & 3000 & 3280 & 3600\\ 
WDJ174435.13$-$725935.67 & 109.234T.001 & 2022-05-31T05:38:18 &  1.521\ldots 1.503 &  0.85\ldots 0.63 & 4080 & 4260 & 1500\\ 
WDJ191429.35$-$544019.71 & 109.234T.001 & 2022-04-10T08:51:23 &  1.232\ldots 1.194 &  1.61\ldots 0.94 & 1600 & 1840 & 1920\\ 
             & 111.255J.001 & 2023-04-17T07:15:57 &  1.390\ldots 1.273 &  0.71\ldots 0.54 & 3000 & 3280 & 3600\\ 
WDJ194919.78$-$271225.74 & 109.234T.001 & 2022-04-27T08:38:30 &  1.063\ldots 1.017 &  1.28\ldots 0.84 & 2460 & 2640 & 2880\\ 
             & 111.255J.001 & 2023-05-14T07:28:10 &  1.068\ldots 1.020 &  0.84\ldots 0.63 & 2600 & 2880 & 3200\\ 
WDJ195219.68$-$384614.05 & 109.234T.001 & 2022-04-27T07:12:34 &  1.295\ldots 1.149 &  1.28\ldots 0.89 & 3180 & 3360 & 3600\\ 
WDJ195347.76$-$430713.92 & 109.234T.001 & 2022-05-02T07:19:53 &  1.233\ldots 1.107 &  1.25\ldots 0.88 & 4080 & 4260 & 1500\\ 
WDJ201730.00$-$171240.89 & 109.234T.001 & 2022-05-12T08:56:03 &  1.019\ldots 1.009 &  1.43\ldots 1.01 & 2460 & 2640 & 2880\\ 
WDJ215453.40$-$302918.67 & 109.234T.001 & 2022-05-27T07:59:55 &  1.116\ldots 1.077 &  0.93\ldots 0.77 & 1590 & 1680 & 1800\\ 
             &              & 2022-06-02T06:28:02 &  1.407\ldots 1.198 &  1.42\ldots 0.87 & 3180 & 3360 & 3600\\ 
\hline
\end{tabular}
\end{table*}

\section{Observations and data processing}\label{sec:obs}
\subsection{Observations}\label{ssec:data}
Low resolution spectroscopy and/or photometric SEDs can, in principle, provide enough information to evaluate how accurately a model can reproduce the observed flux of a star. However, several effects (e.g.  contamination from neighbours or background sources at specific wavelengths, presence of magnetic fields or of weak lines) 
that would negatively effect the robustness of the target as a flux standard would not be visible in these limited datasets.
The most reliable way to verify the predicting power of our models and test the potential of star as a flux standard, is to acquire middle to high resolution spectra covering the entire wavelength range of interest. 
Having the full spectra is a final proof of the correctness of the models.
The 24 southern DA white dwarf candidates cover the full range of RA for observations with the X-shooter instrument \citep{vede11} on the VLT UT3. X-shooter is a medium resolution echelle spectrograph, whose three arms cover the wavelength range 300--550\,nm (UVB, $R$\,=\,3000--10\,000), 550--1000\,nm (VIS, $R$\,=\,5000--18\,000), and 1000--2480\,nm (NIR, $R$\,=\,4000--11\,500). The instrument uses slits of various widths (0.5\arcsec--5.0\arcsec) with a length of 11\arcsec. For our observations, we used the widest slit of 5\arcsec\ to collect the flux as completely as possible, and the NODDING mode to improve sky subtraction. 

Observational information on the initial set of 24 flux standard candidates is summarised in Table~\ref{tab:obs_info}. For each target the ESO programme number (`Run') is given, the date and time at the start of the observations, the range covered in airmass and of the seeing (at 500\,nm) at zenith during the observations, and total exposure times in the UVB, VIS, and NIR arms of X-shooter.

\begin{figure*}[ht]
\begin{center}
\includegraphics[width=.85\linewidth]{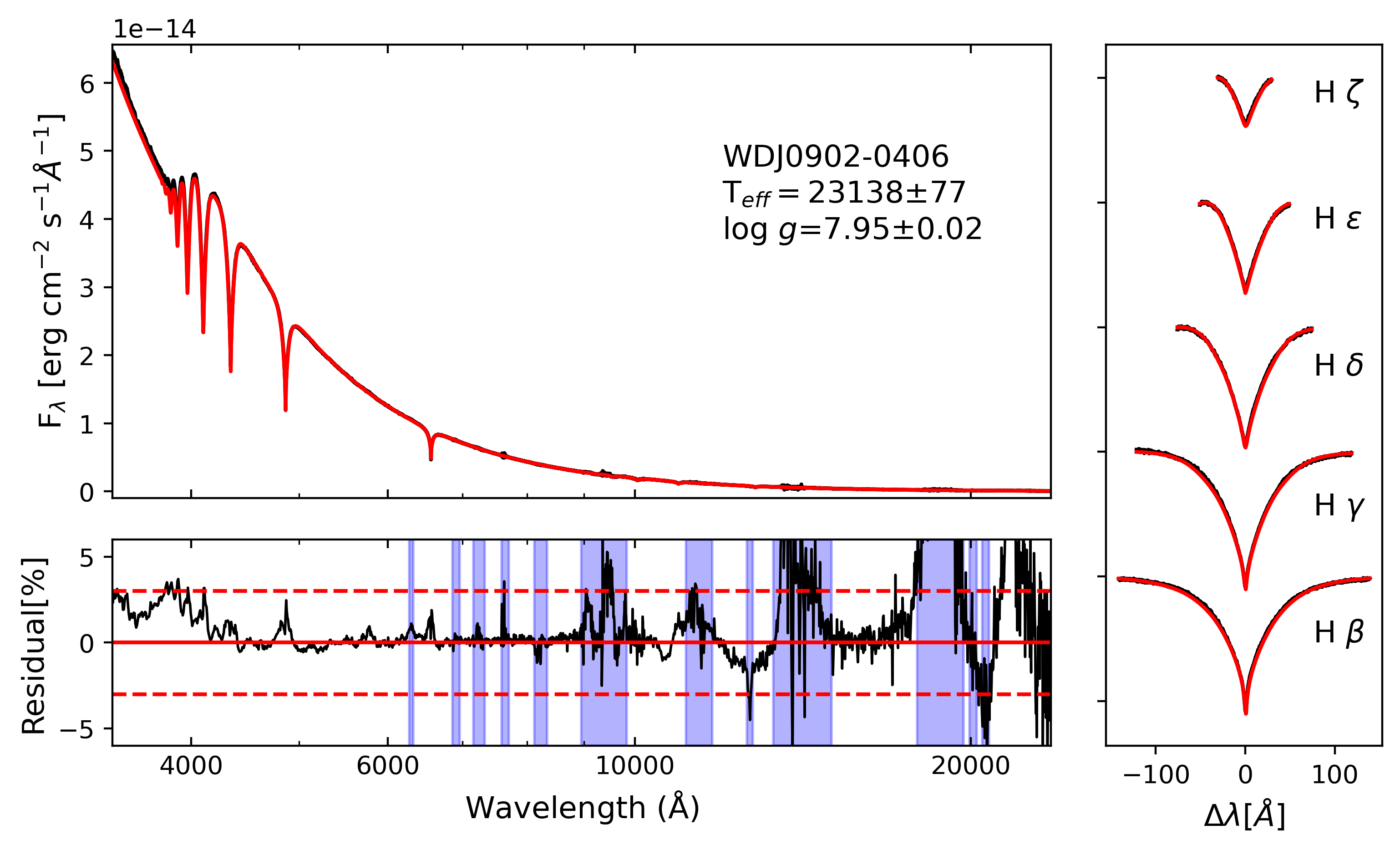}
\caption[]{Spectrophotometric fits of the X-shooter data for our sample white dwarf WDJ0902$-$0406. \textit{Top-left panel}: SED fit between the observed spectrophotometry (black) and best-fitting model (red). The adopted atmospheric parameters are indicated. \textit{Bottom-left panel}: Flux residuals from the corresponding SED fit. The black line
is the calculated residual, and red horizontal lines show residuals of 0 and $\pm$3\% as a
guide. The vertical blue bands indicate wavelength ranges contaminated by telluric lines. \textit{Right panel}: Balmer line fits for H$\beta$ to H$\zeta$ between the observed
spectrophotometry (black) and best-fitting model (red). The line profiles are vertically offset for clarity. Analogous plots for the remaining 13 flux standard white dwarfs are shown in Appendix~\ref{appendix:A}.}
\label{fig:sample_fit}
\end{center}
\end{figure*}

\subsection{Stray light problems}\label{ssec:stray}
Using wide slits proved effective in the UVB and VIS arms, yielding a higher signal-to-noise ratio (S/N) than predicted by the Exposure Time Calculator (ETC), likely due to better seeing, lower water vapour, and better transparency (clear instead of thin cirrus). However, the NIR arm data showed much lower S/Ns than the predictions of the ETC. Detailed checks of the raw  NIR data showed that the observed object flux in regions free of sky emission lines was about 20\% to 40\% higher than predicted, which was consistent with the results in the UVB and VIS arms. The NIR background level between sky emission lines, however, was about twice as high as predicted by the ETC. Stray light from the $K$-band was the most likely culprit for the additional background. A higher background level causes higher noise, which will be preserved in the sky- and background-corrected extracted spectra. Because our targets are faint in the NIR, the higher background reduces the S/N by a factor of about 2--4 compared to the ETC predictions (see Fig.\,\ref{fig:NIR_SNR_ratio}). For P112 we therefore requested a slit of 1.2\arcsec\,$\times$\,11\arcsec\ for the NIR arm. As can be seen in Fig.\,\ref{fig:NIR_SNR_ratio} the discrepancy between the predicted and observed S/N has been substantially reduced.

\subsection{Data reduction}\label{ssec:data_reduction}
All raw spectra were processed using the ESO X-shooter pipeline (version xshoo-3.6.3)
together with the EsoReflex workflow \citep{Reflex}. We changed the recipe parameter {\tt stdextract-interp-hsize} of the recipe {\tt xsh\_scired\_slit\_nod} from 30 to 150 to avoid the spurious region of zero flux in the VIS data. It turned out that this choice also improved the quality of some of the NIR spectra. We note that the pipeline uses the MOLECFIT recipe \citep{Smetteetal15,Kauschetal15} for telluric line correction.

During the analysis, we found the use of new reference data for the {X-shooter} flux standard stars \citep{Sanaetal24}
improved the agreement between the atmospheric parameters derived from photometric data and from flux-calibrated spectra. Therefore, the {X-shooter} data were reduced again, using the new reference data for the determination of the response. This selection excluded the use of the standard star \object{Feige 110}, which shows prominent helium absorption that is not included in the new reference data.

In some cases, the spectrophotometric standard star observed closest in time to our data provided an imperfect response, most likely due to the faintness of the standard star (\object{GD 71} or \object{GD 153}). In these cases, we selected another standard star up to one month away that provided a good response. Because the instrumental response varies very slowly, this is not a problem.

\section{Modelling of the observed spectra}\label{ssec:model_spectra}
All but one of the observed targets were confirmed as DA white dwarfs and kept in the sample for further characterisation. 
To determine the \Teff\ and $\log g$ for our stars, we used the so-called spectroscopic method, which compares the observed continuum-normalised Balmer line profiles of a DA white dwarf with an appropriate set of model spectra. Following this method the stellar parameters of a white dwarf can be determined from a spectrum largely independently of its flux calibration.
In our analysis we adopted the latest update (March 2024) of the grid of 3D pure-hydrogen (DA) local thermal equilibrium (LTE) atmosphere models described in \citet{Tremblayetal13} and \citet{Tremblay2015a}. The grid covers the wavelength range 4 to 60\,000\,nm and includes models for 7.0\,$\leq$\,$\log g$\,$\leq$\,9.0 in steps of 0.5\,dex (cgs units) and 1500\,$\leq$\,$\Teff$\,$\leq$\,40\,000\,K. 
Above $40\,000$\,K deviations from LTE (so-called NLTE effects) start to become noticeable \citep{Napi2020,Munday2024}. Hence, for the hottest star in the sample, we used an appropriate NLTE grid covering the range 5000\,K to 140\,000\,K in $T_\mathrm{eff}$ \citep{Tremblay2011}.
 
All details about the models' input physics and numerical methods are available on P.-E. Tremblay's Source Model Data web page\footnote{\url{https://warwick.ac.uk/fac/sci/physics/research/astro/people/trem)blay/modelgrids/}}. 
In our fitting routine, principal component analysis methods within the \texttt{python scikit-learn} package were used to generate arbitrary models between grid nodes. 

\begin{figure}[ht]
\centering
\includegraphics[width=.98\linewidth]{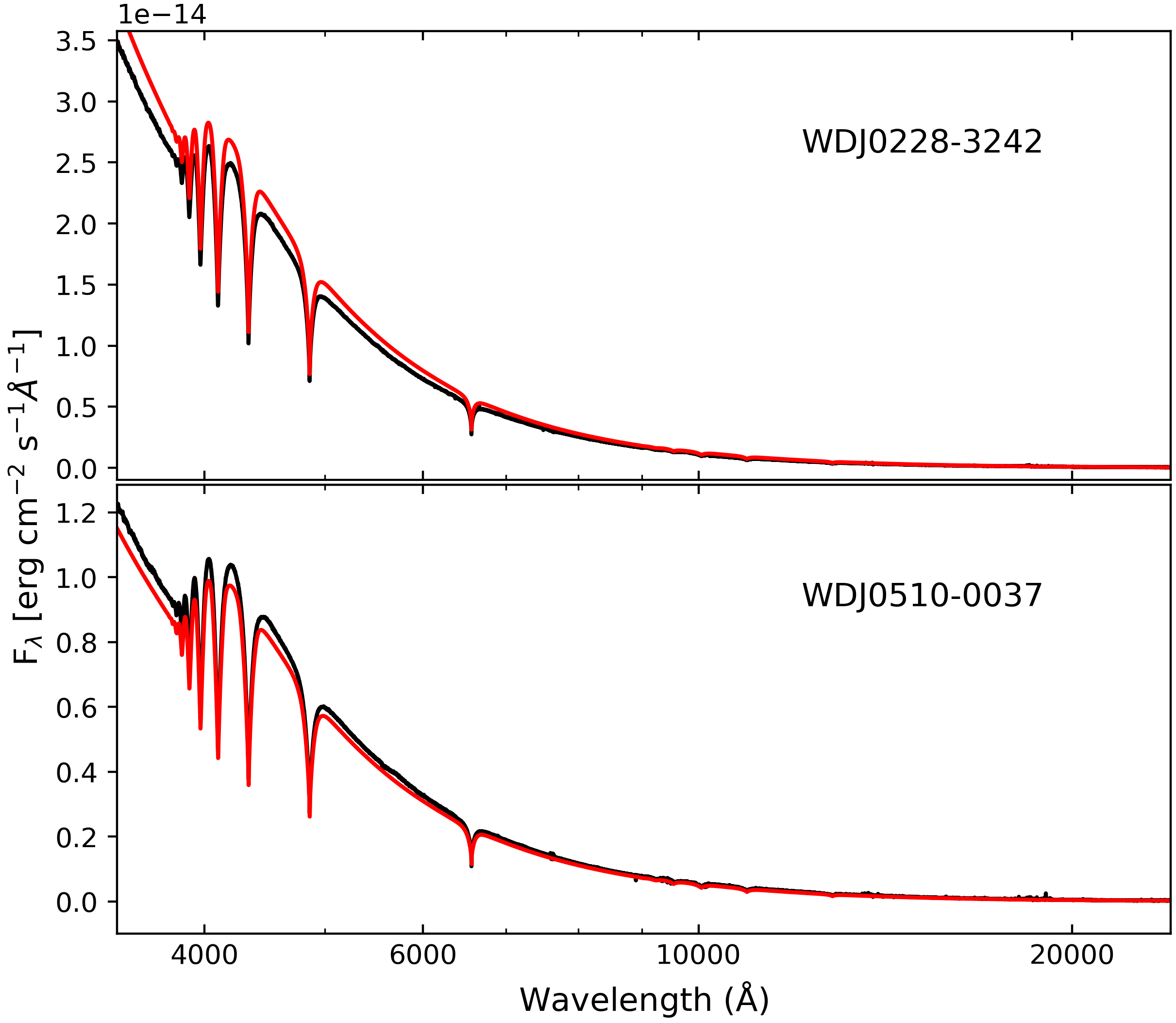}
\caption[]{Examples of observed flux offsets in some exposures. Model fluxes (red lines) are compared with the pipeline-reduced and pre-finally calibrated observed fluxes (black line) for two of the flux standards.}
\label{fig:abs_flux_offset}
\end{figure}

Before comparing the models to the observed spectra, the arbitrarily high natural resolution of the models needed to be degraded to match that of X-shooter. We achieved this by convolving the models in logarithmic wavelength space with a Gaussian function with full-width-at-half-maximum equal to the reciprocal of the nominal resolving power $R$ of the X-shooter UVB arm.
Both the model spectra and the X-shooter UVB spectra were continuum normalised by fitting a spline through the continuum regions (see e.g. the top-left panel of Fig.~\ref{fig:sample_fit}). 
The cropped H$\beta$ to H$\zeta$ absorption features were then fitted with our DA model spectra adopting a $\chi^2$ minimisation routine using the \texttt{scipy optimize} trust region reflective (\texttt{trf}) algorithm, which is a non-linear least-squares method \citep{Byrd1987}. \Teff, $\log g$ and an arbitrary wavelength shift were free parameters in this fitting routine. The wavelength shifts that achieve the best fit to the Balmer lines largely depend on the shape and size of the wings of the lines (which in turn depend on \Teff\ and $\log g$) and do not necessarily provide the best match for the line cores that only have a low weight in the fit 
(see e.g. the right panel of Fig.~\ref{fig:sample_fit}). Consequently, though these wavelength shifts were included in the fits, they do not constitute reliable measurement of the radial velocity of our targets and we do not report them.
We also opted to exclude the H$\alpha$ Balmer line from the fits, as this is covered by the VIS arm of the spectrograph. Inclusion of this line would, therefore, have required both degrading this region of the models to a different resolution and stitching UVB and VIS spectra together. 
Our tests show that these additional steps can negatively impact the continuum normalisation and the overall quality of the fit.

\begin{figure*}[t]
\begin{center}
\includegraphics[width=1.0\linewidth]{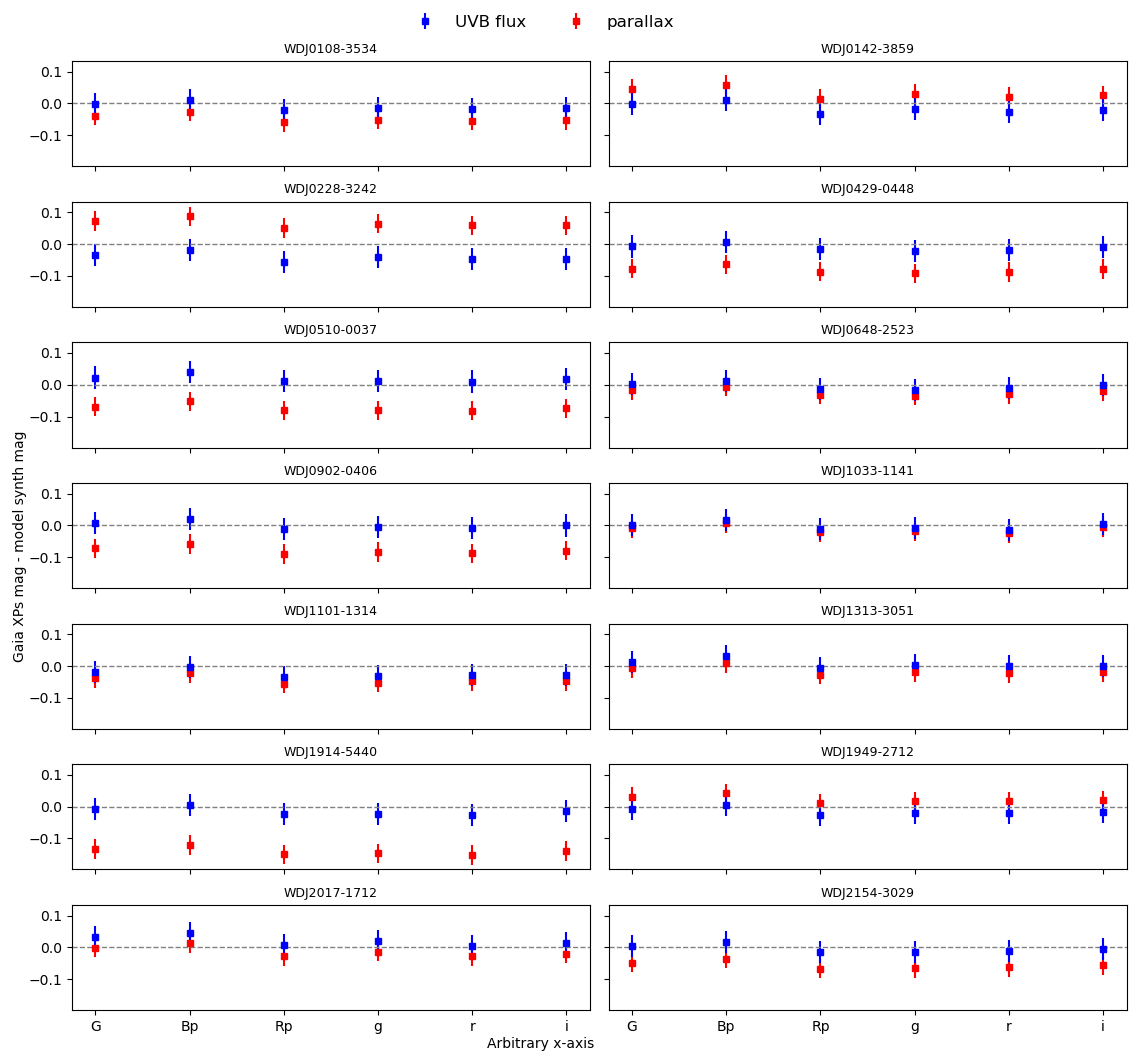}
\caption[]{Difference between synthetic \textit{G, Bp, Rp, g, r}, and \textit{i} magnitudes derived from models scaled to match the UVB flux (blue) and scaled to the \textit{Gaia} parallax  (red) with the observed \textit{Gaia} \textit{G, Bp}, and \textit{Rp} magnitudes, and with the \textit{g, r}, and \textit{i} magnitudes calculated from the \textit{Gaia} XP spectra \citep{montegriffoetal23-1}, for the 14 white dwarfs in our final sample. The size of the error bars depends largely on the propagation of the uncertainties in the stellar parameters corresponding to the models.}
\label{fig:synth_comp}
\end{center}
\end{figure*}

\begin{figure}[]
\begin{center}
\includegraphics[width=.995\columnwidth]{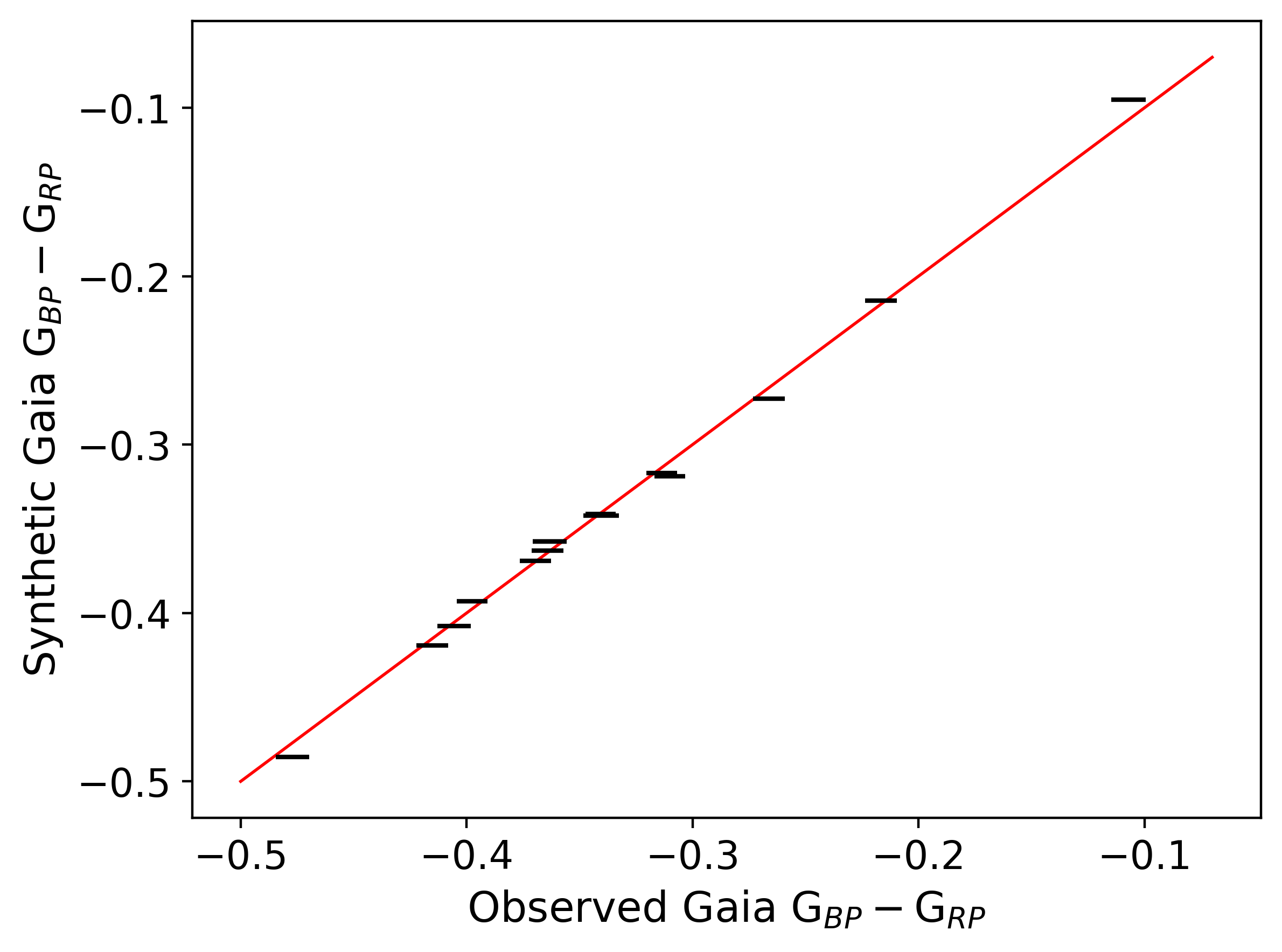}
\caption[]{Comparison of \textit{Gaia} colours. Observed \textit{Gaia} $Bp-Rp$ colours of the flux standards are compared with synthetic values derived from the model fluxes. The red line marks the 1:1 relation.}
\label{fig:colours}
\end{center}
\end{figure}

Once the spectroscopic best-fitting parameters (\Teff\ and $\log g$) were determined, the model flux was scaled to the white dwarf's distance (obtained from the \textit{Gaia} parallax) and radius. The latter was calculated by adopting the DA $M$-$R$ relation with thick hydrogen layers and carbon-oxygen cores from  \citet{Bedard2020}\footnote{\url{https://www.astro.umontreal.ca/~bergeron/CoolingModels/}} using the best-fitting \Teff\ and $\log g$. An example of the residuals from the comparison of the observed and model fluxes is shown in the bottom-left panel of Fig.~\ref{fig:sample_fit}. The largest deviations occur in regions where the telluric line removal is not perfect due to saturated telluric lines. Further systematic deviations occur in the UV wavelength region below 340\,nm due to the Huggins bands of atmospheric O$_3$ in some stars.
Analogous plots to Fig.~\ref{fig:sample_fit} for the other flux standard stars can be found in Fig.~\ref{fig:sample_fit_app}.

\section{Evaluation of the flux standard candidates}\label{sec:calib}
As a first check to evaluate the suitability of our targets as flux standards, we compared the spectroscopic \Teff\ and $\log g$ values from our fits with the corresponding photometric parameters published in \citet{GentileFusilloetal21}. Differences in these sets of parameters are potentially indicative of physical  properties that can render the star unsuitable as a standard (e.g. binarity, variability, etc.). We excluded from our final sample all targets that exhibit a discrepancy larger than 3$\sigma$ in their parameters. It is worth noting that this is a very conservative approach and this level of discrepancies between photometric and spectroscopic parameters can be expected for individual objects even if there is nothing atypical in either dataset or in the star itself. In other words, the stars we excluded only on the basis of this criterion are not necessarily unsuitable as flux standards, but further test would be required to exclude potential issues. 

\begin{table*}[ht]
\centering
\caption[]{Spectrophotometric standard stars for MICADO.}
\label{tab:sample}
\setlength{\tabcolsep}{1.7mm}
\begin{tabular}{lllrrrrrr}
\hline\hline
Object name & $K$ & $G_{RP}$ & RA(J2000) & Dec.(J2000) & PM$_{R.A.}$  & PM$_{Dec}$ & $T_\mathrm{eff}$  & $\log g$\\
\cline{2-3} \cline{6-7}
\& \textit{Gaia} DR3 ID & \multicolumn{2}{c}{[Vegamag]} & [h m s] & [$^\circ$ ' ''] & \multicolumn{2}{c}{[mas\,yr$^{-1}$]} & 
[K] & [cm\,s$^{-2}$]\\
\hline
WDJ010820.77$-$353441.70 & 15.86$^{1}$  & 14.98 & 01 08 20.77 & $-$35 34 41.70 & $-$38.560 &  $-$52.062 &  29095$\pm$89 & 7.94$\pm$0.01 \\
5014009353235167744 \smallskip
\\
WDJ014251.02$-$385906.76 & 15.01$^{1}$ & 14.56 & 01 42 51.02 & $-$38 59 06.76 &   129.763 &     75.890 & 21862$\pm$75 & 7.97$\pm$0.01 \\
4962193390308361728\smallskip
\\
WDJ022827.20$-$324233.80 & 14.70$^{1}$ & 13.75 & 02 28 27.20 & $-$32 42 33.80 & $-$69.882 & $-$217.808 & 22537$\pm$84 & 7.99$\pm$0.01 \\
5063539946887524864\smallskip
\\
WDJ042926.24$-$044845.62 & 14.80$^{1}$ & 14.58 & 04 29 26.24 & $-$04 48 45.62 & $-$58.709 &     61.075 & 14000$\pm$21 & 8.08$\pm$0.02\\
3202808828330265088\smallskip
\\
WDJ051002.85$-$003755.65 & 15.37$^{2}$ & 14.97 & 05 10 02.85 & $-$00 37 55.65 & $-$7.644  &     30.287 & 17691$\pm$31 & 7.96$\pm$0.02 \\
3227446822324601472\smallskip
\\
WDJ064856.09$-$252346.96 & 14.53$^{1}$ & 13.90 & 06 48 56.09 & $-$25 23 46.96 & $-$25.355 &     34.256 & 27557$\pm$23 & 7.93$\pm$0.01\\
2921786919133198848\smallskip
\\
WDJ090217.30$-$040655.48 & 13.90$^{1}$ & 13.40 & 09 02 17.30 & $-$04 06 55.48 &  $-$4.334 &      6.399 & 23138$\pm$77& 7.95$\pm$0.02 \\
5762406957886626816\smallskip
\\
WDJ103342.76$-$114138.34 & 13.72$^{2}$ &13.25 & 10 33 42.76 & $-$11 41 38.34 &$-$343.949 &  $-$27.989 &  25644$\pm$52 & 7.96$\pm$0.01\\
3754712881779364992\smallskip
\\
WDJ110112.24$-$131442.23 & 15.44$^{1}$& 15.16 & 11 01 12.24 & $-$13 14 42.23 &     0.763 &   $-$6.923 &  25030$\pm$103 & 8.73$\pm$0.03\\

3564376149017654272\smallskip
\\
WDJ131341.58$-$305133.71 & 14.91$^{1}$ &14.70 & 13 13 41.58 & $-$30 51 33.71 &    25.565 &  $-$73.355 & 20105$\pm$65 & 8.00$\pm$0.02\\
6182345731690327808\smallskip
\\
WDJ191429.35$-$544019.71 & 15.18$^{1}$ & 14.89 & 19 14 29.35 & $-$54 40 19.71 &    21.246 &  $-$36.620 & 25406$\pm$75 & 8.38$\pm$0.03\\
6643352368598722944\smallskip
\\
WDJ194919.78$-$271225.74 & 14.90$^{1}$ & 14.41 & 19 49 
19.78 & $-$27 12 25.74 &    50.521 & $-$164.678 & 25873$\pm$27 & 7.79$\pm$0.02\\
6754497811462316416\smallskip
\\
WDJ201730.00$-$171240.89 & 15.16$^{1}$ & 14.59 & 20 17 30.00 & $-$17 12 40.88 & $-$64.125 & $-$119.410 & 41784$\pm$146 & 7.93$\pm$0.02\\
6873359343451933312\smallskip
\\
WDJ215453.40-302918.67& 15.33$^{1}$ & 14.97 & 21 54 53.40 & $-$30 29 18.67 &    26.231 &  $-$13.517 & 28715$\pm$109 & 8.32$\pm$0.02\\
6616313457820826496\smallskip
\\
\hline

\end{tabular}
\tablefoot{$^{1}$ $K$-band magnitude from 2MASS \citep{Cutrietal03}, 
$^{2}$ $Ks$-band magnitude from VHS \citep{McMahonetal21}, coordinates and proper motions from the \href{https://gea.esac.esa.int/archive/}{Gaia Data Release 3} corresponding to epoch 2016.}
\end{table*}

In theory, once scaled to the correct distance, the best-fitting models for the observed white dwarfs can be compared directly with the full X-shooter spectra (UVB+VIS+NIR) to assess any relative difference between the two.
However, for half of the targets, the total flux of the \textit{Gaia} parallax scaled model showed an offset compared to the observed X-shooter flux, under-predicting or over-predicting observations by up to 14\% (see Fig.~\ref{fig:abs_flux_offset} for examples).
To determine which one provides a better absolute flux reference, we created a second set of scaled models by multiplying our model fluxes by an arbitrary scaling factor so that it matches that of the observed X-shooter spectrum over the wavelength range 450--460\,nm (a featureless region in the white dwarf spectrum in the UVB arm). We then calculated synthetic \textit{G, Bp, Rp, g, r}, and \textit{i} magnitudes from both sets of models (UVB scaled and \textit{Gaia} parallax scaled) and compared them with the observed \textit{Gaia} \textit{G, Bp}, and \textit{Rp} magnitudes and with the \textit{g, r}, and \textit{i} synthetic magnitudes calculated from the \textit{Gaia} low-resolution, externally calibrated \textit{Bp, Rp} spectra (XP spectra, \citealt{montegriffoetal23-1}).

Figure~\ref{fig:synth_comp} clearly shows that the magnitudes derived from the UVB scaled models are in better agreement with the observed magnitudes. Hence, we adopted this UVB scaling as the absolute flux of our reference models. This finding further justifies our decision to observe using a 5.0\arcsec slit and shows that this did not result in any significant flux loss in the UVB observations.
However, our P112 NIR observations were carried out with a 1.2\arcsec\,$\times$\,11\arcsec\ aperture in order to improve background subtraction in data reduction. This results in a marked step-down between the flux observed in UVB and VIS, and the flux observed in the NIR. Consequently, in order to compare the observed NIR flux with with the models, we had to first scale it up (by up to 50\%) to align with the UVB and VIS arms and so match our models.

After correcting the spectra for these absolute flux offsets, we compared them with the best-fitting models and examined the residual, effectively probing how accurately our models predict the full SED of the white dwarfs (e.g. Fig.~\ref{fig:sample_fit}). Targets that displayed a slope in the residual, and hence a colour-dependent offset, exceeding 3\% were excluded from the sample. 
In the end, out of the 24 observed targets, we selected 14 reliable flux calibrators (see Table~\ref{tab:rejected} for additional information on the rejected targets).  

As an additional test, we computed synthetic G$_{BP}$ and G$_{RP}$ magnitudes from our adopted model fluxes for our flux standard stars and compared the model G$_{BP}-$G$_{RP}$ colours with the observed \textit{Gaia} ones. Excellent agreement was found, as shown in Fig.~\ref{fig:colours}. 

Information on the final 14 faint southern flux standard stars for MICADO is summarised in Table~\ref{tab:sample}. For each object, we report the $K$-band and \textit{Gaia} $G_{RP}$ magnitudes (the latter indicates the brightness in the bandpass where the wavefront sensor of the MICADO SCAO module will operate), positional information in RA and Dec., proper motions $PM_{RA}$ and $PM_{Dec}$, and the values of \Teff\ and $\log g$ determined in this study. 
In Table~\ref{tab:rejected} we list the target white dwarfs that were observed with X-shooter but were rejected as potential flux standards, with the reasons for the rejection reported in the final column.

\begin{figure}[ht]
\includegraphics[width=.995\columnwidth]{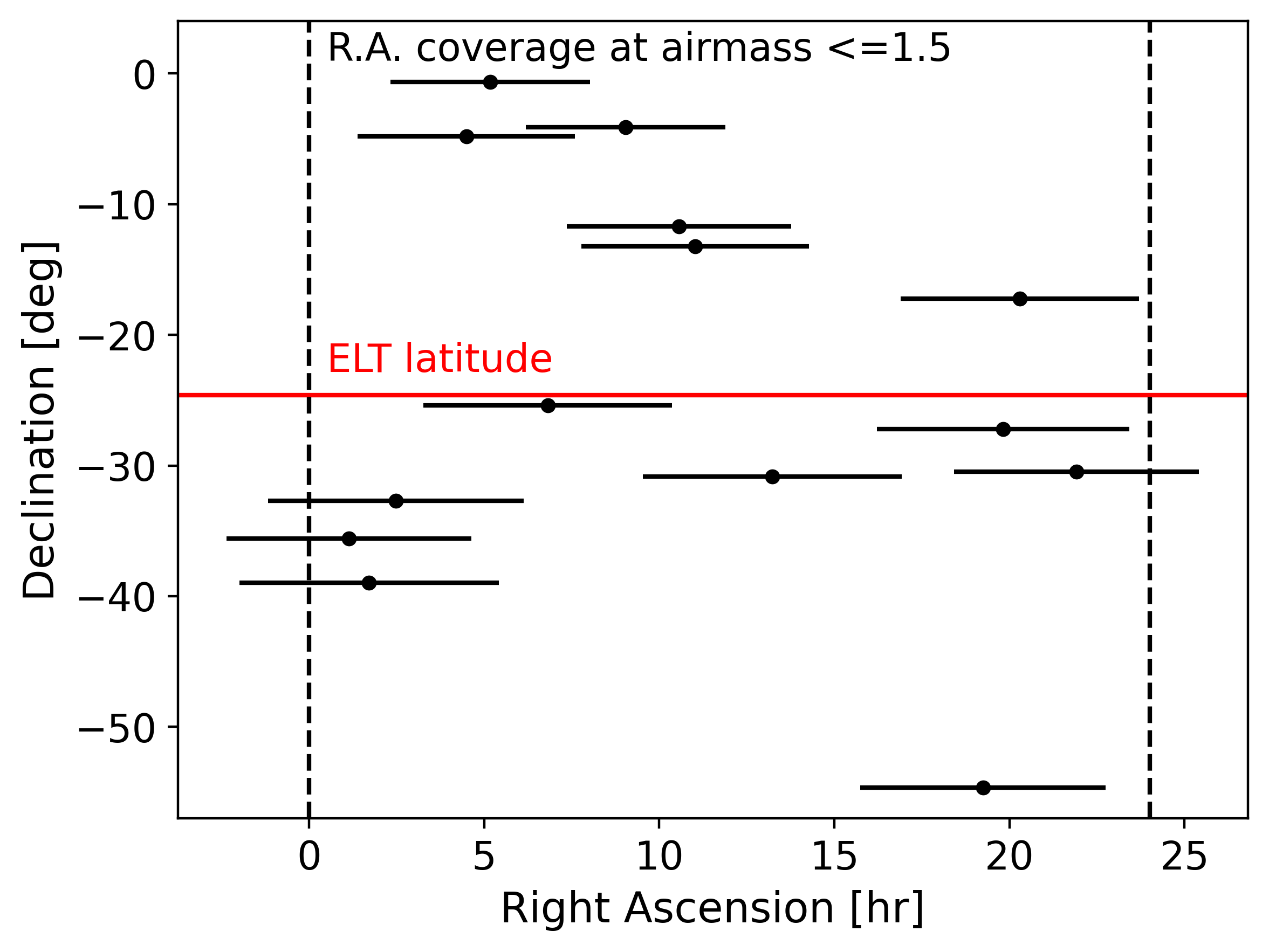}\\[-5mm]
\caption[]{RA and Dec. distribution of the 14 standard stars from Table~\ref{tab:sample} and their RA coverage at an airmass of up to 1.5. The red line marks the latitude of the ELT, illustrating that the range in RA is fully covered by stars south of the ELT.
 \label{fig:std_mic}}
\end{figure}

The distribution of the flux standard stars from Table~\ref{tab:sample} in RA and Dec. is shown in Fig.~\ref{fig:std_mic}, with the ranges indicating coverage in RA at an airmass of up to 1.5. The subset of the proposed flux standard stars south of the ELT latitude alone is sufficient to provide full sky coverage, which is important for occasional conditions at the ESO ELT and Paranal sites with prevalent wind from the north.   

However, for the use of our flux standard stars with MICADO there will be some additional constraints.
Initially, MICADO will be operated in a stand-alone phase where only SCAO will be available; MCAO will be realised only some years later via the Multi-conjugate adaptive Optics Relay For ELT Observations \citep[MORFEO;][]{Ciliegietal24}. The performance of the AO correction will depend on wavelength, the brightness of the NGSs and the quality of the atmospheric conditions.

As our flux standard stars have been chosen not to have any other objects of similar brightness nearby, they will have to act themselves as NGSs for observations with MICADO in the initial SCAO stand-alone phase. Under median atmospheric conditions the AO performance will drop towards zero for NGSs at $G_{RP}$ about 16\,mag in the $HK$ bands and at about 15.25\,mag in the $IzJ$ bands. Our flux standards will fulfil these conditions except for observations in the $IzJ$ bands under poor atmospheric conditions, where the achievable Strehl ratios will be limited to a few percent. However, in these cases operations are expected to focus on longer wavelengths. For MCAO three NGSs with $H$\,$<$\,21\,mag and $R$\,$<$\,23\,mag will be required in the patrol field of MORFEO around the MICADO field of view, but more specific criteria have not been assessed yet and the necessary star catalogues have not yet been compiled. On the basis of the available photometry, most of our targets should be suitable for operation, but a formal feasibility test for the use of the complete sample of our flux standard stars with MCAO will have to come at a later date. 

\section{Conclusions}\label{sec:conclusions} 

We have identified and characterised 14 faint southern DA white dwarfs to be used as flux reference standards for NIR observations, with a specific focus on the first generation ELT instrument MICADO. We obtained X-shooter spectroscopy covering the wavelength range 340--2480\,nm and fitted the observed Balmer line profile using the pure-hydrogen atmosphere models described in \citet{Tremblayetal13} and \citet{Tremblay2015a} to obtain \Teff\ and $\log g$. We then used \textit{Gaia} DR3 parallax measurements and the DA $M$-$R$ relation \citep{Bedard2020} to scale the best-fitting models to the absolute observable flux.  
We find that the observational limitations of X-shooter for NIR-faint targets and non-ideal flux calibration cause significant offsets in the absolute flux between our distance-scaled white dwarf models and the X-shooter spectra, particularly in the NIR region. 
Nonetheless, this work shows that, after scaling the models to match the observed UVB flux and correcting for the absolute flux offset in the NIR, our models predict the SED of the selected white dwarfs to better than 3\% across the entire wavelength range of interest.
Furthermore, \Teff\ and $\log g$ values obtained from our model fits agree within 1.5$\sigma$ with the results of photometric fits based on observed \textit{Gaia} EDR3 photometry \citep{GentileFusilloetal21}, demonstrating good agreement between independent determinations of atmospheric parameters for stars in our sample.
As an additional test, we also computed synthetic \textit{Gaia} $Bp-Rp$ colours from our reference models and compared them with the observed ones, finding excellent agreement for the entire sample. 

To conclude, we have established a network of 14 white dwarf flux standards in the magnitude range 13.9 to 15.9 in the $K$ band. The sample uniformly spans the entire RA range and covers declinations between $-$54\degr\ and 0\degr. The stars were selected to serve as flux calibrators for MICADO, but they can also be used for any observatory and as an addition to CALSPEC \citep{bohlinetal2020}, which contains a number of white dwarfs in the same magnitude range, and other modern networks of flux standards \citep{Narayan2016, Narayan2019, Axelrod2023, elmsetal24}. 
Once absolute calibrations of JWST spectra in the NIR become available, we will be able to use X-shooter observations of the primary HST flux standards GD\,71 and GD\,153 to reliably join our new network of flux standards for ESO IR instruments with the HST flux scale. 

\begin{acknowledgements}
We thank E. Valenti for valuable comments on the manu\-script.
NP and WK acknowledge funding for ELT instrumentation projects via the Vice Rectorate for Research
of the University of Innsbruck. This work has made use of data from the European Space Agency (ESA) mission
{\it Gaia} (\url{https://www.cosmos.esa.int/gaia}), processed by the {\it Gaia}
Data Processing and Analysis Consortium (DPAC,
\url{https://www.cosmos.esa.int/web/gaia/dpac/consortium}). Funding for the DPAC
has been provided by national institutions, in particular the institutions
participating in the {\it Gaia} Multilateral Agreement.
We thank the anonymous referee for their constructive report. 
\end{acknowledgements}

\bibliographystyle{aa}
\bibliography{biblio.bib}

\onecolumn
\begin{appendix}
\section{Additional figures}\label{appendix:A}
\begin{figure*}[th]
\begin{center}
\includegraphics[width=.99\columnwidth]{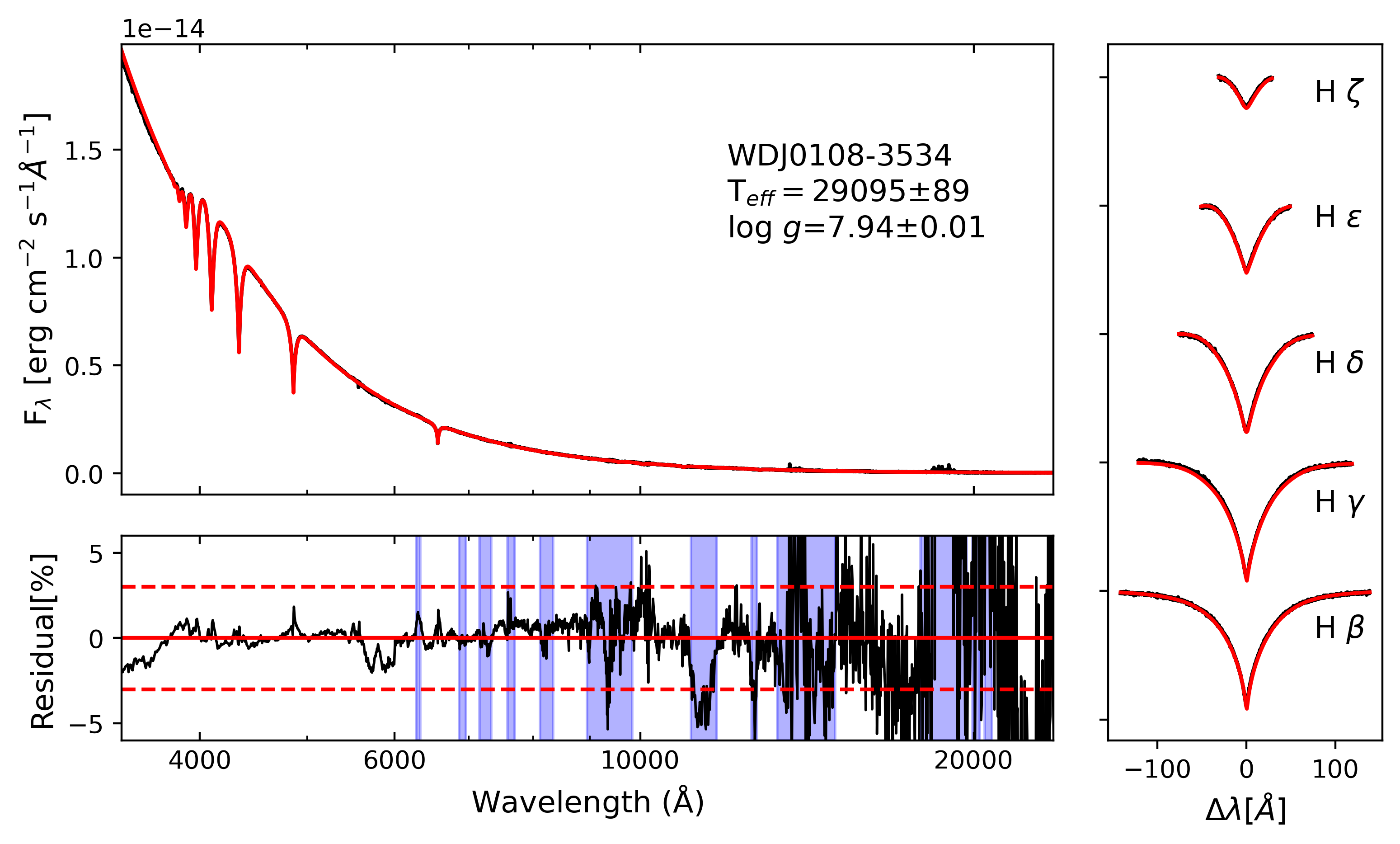}
\includegraphics[width=.99\columnwidth]{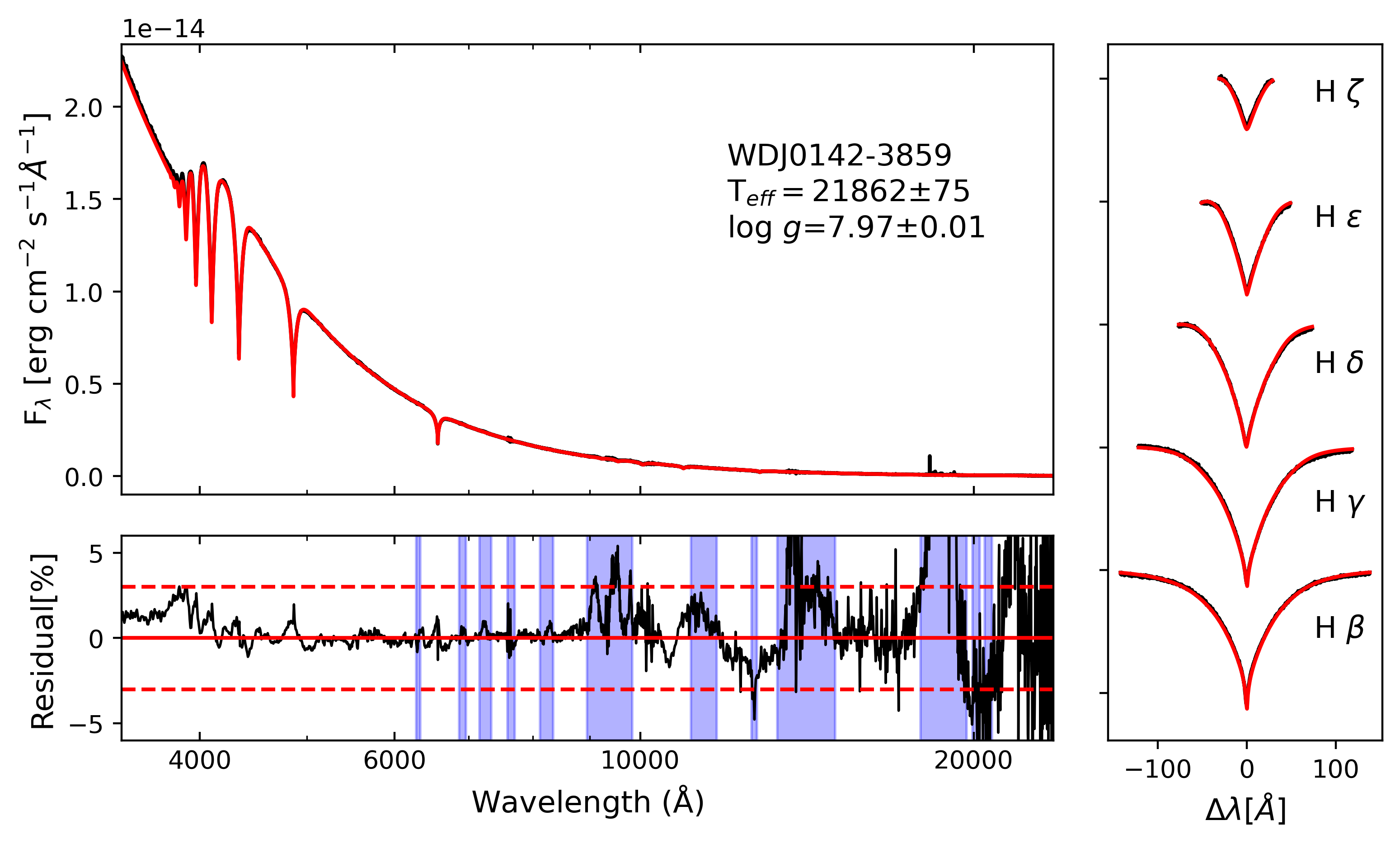}
\caption{Same as Fig.~\ref{fig:sample_fit} but for additional flux standards from the present work.}\label{fig:sample_fit_app}
\end{center}
\end{figure*}

\setcounter{figure}{0}
\begin{figure*}[th]
\begin{center}
\includegraphics[width=.99\columnwidth]{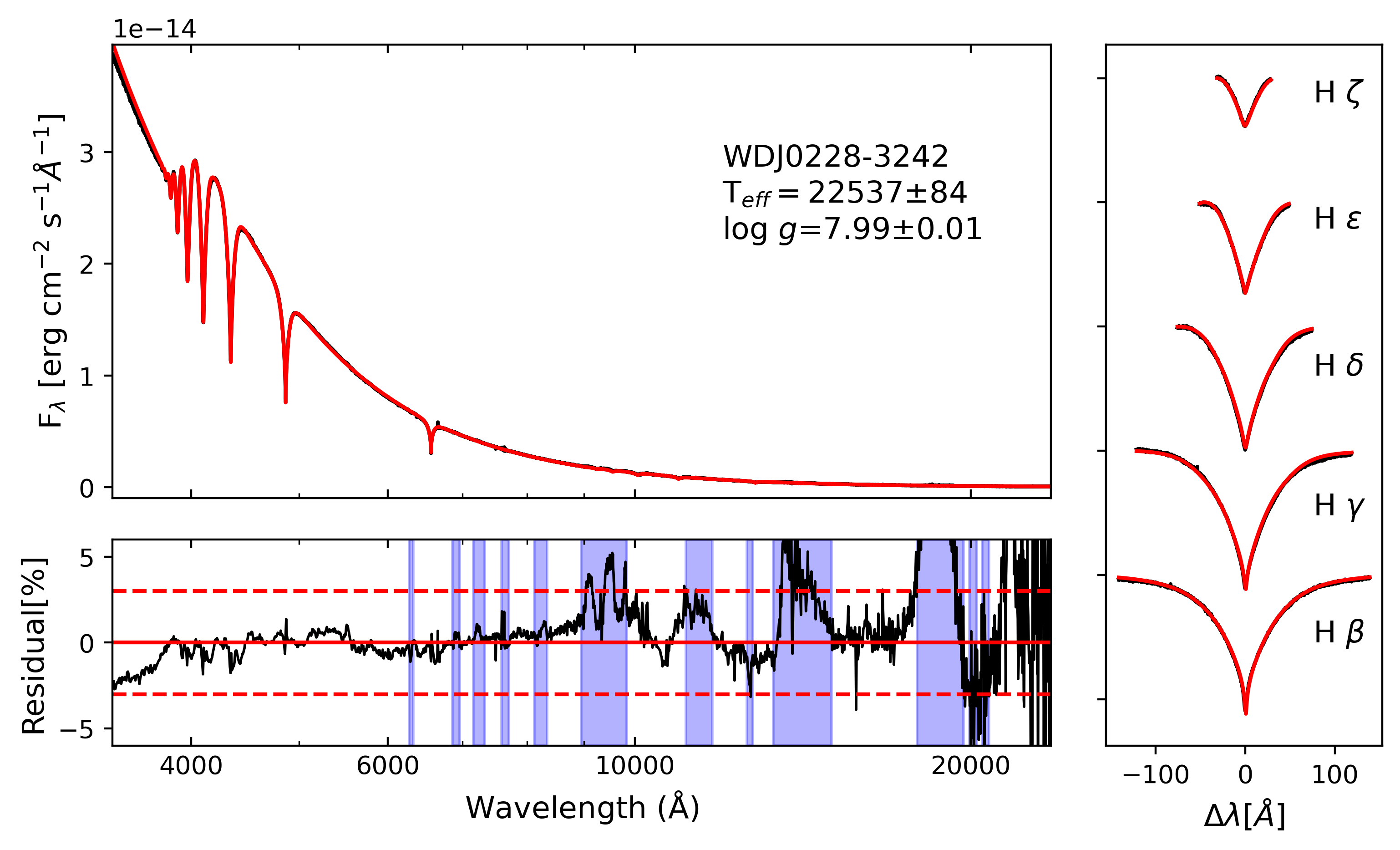}
\includegraphics[width=.99\columnwidth]{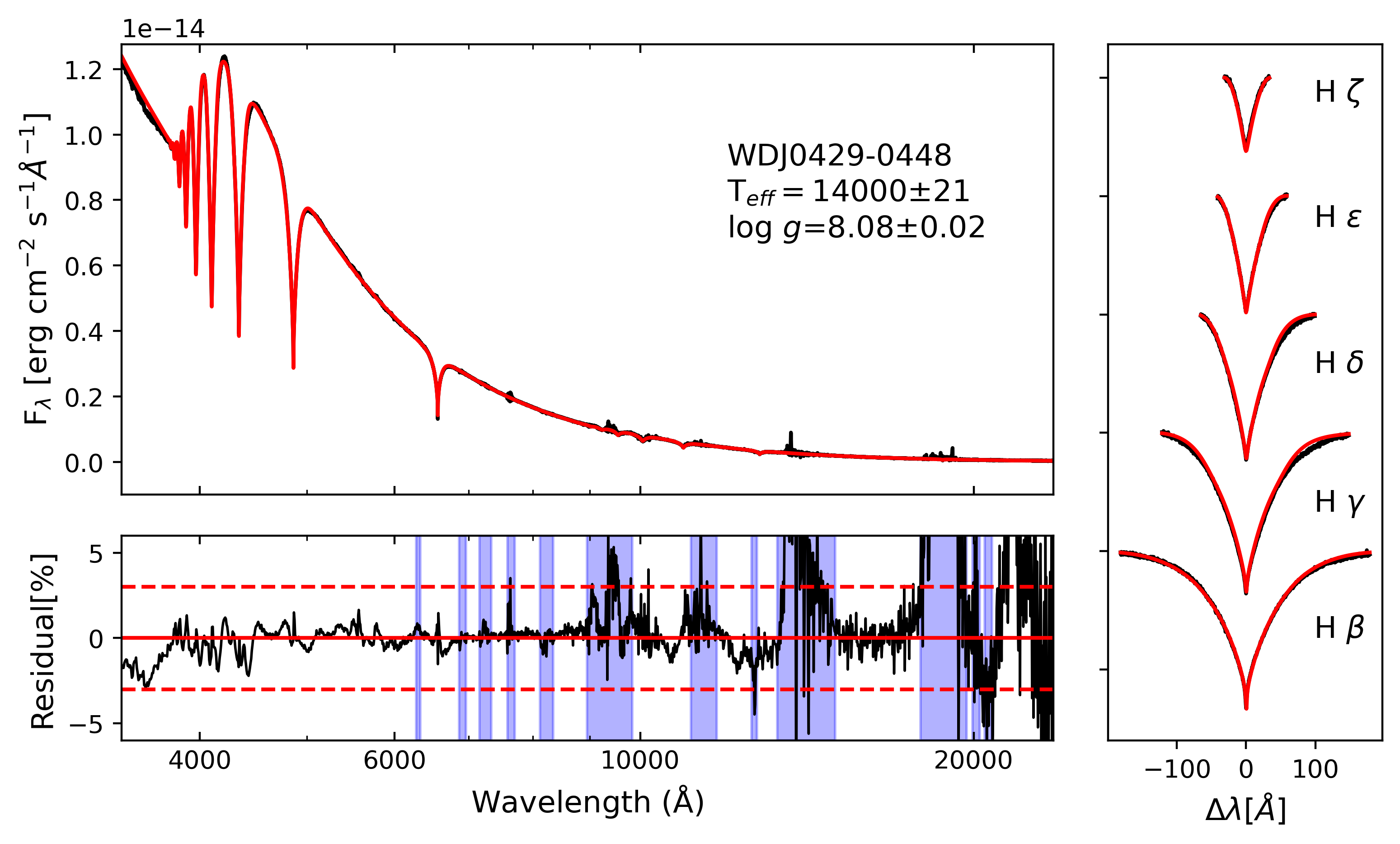}
\caption{Continued.}
\end{center}
\end{figure*}

\setcounter{figure}{0}
\begin{figure*}[th]
\begin{center}
\includegraphics[width=.99\columnwidth]{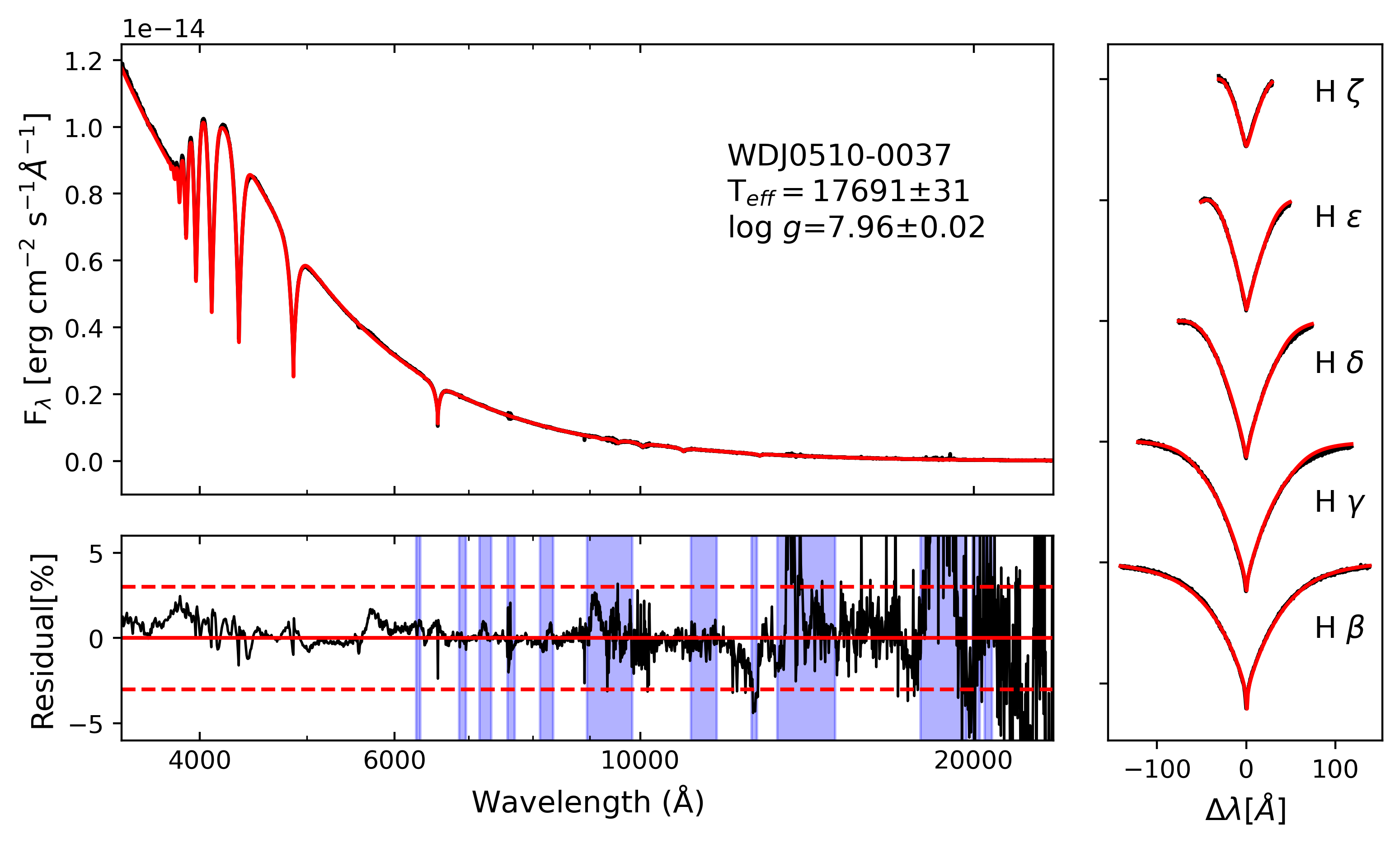}
\includegraphics[width=.99\columnwidth]{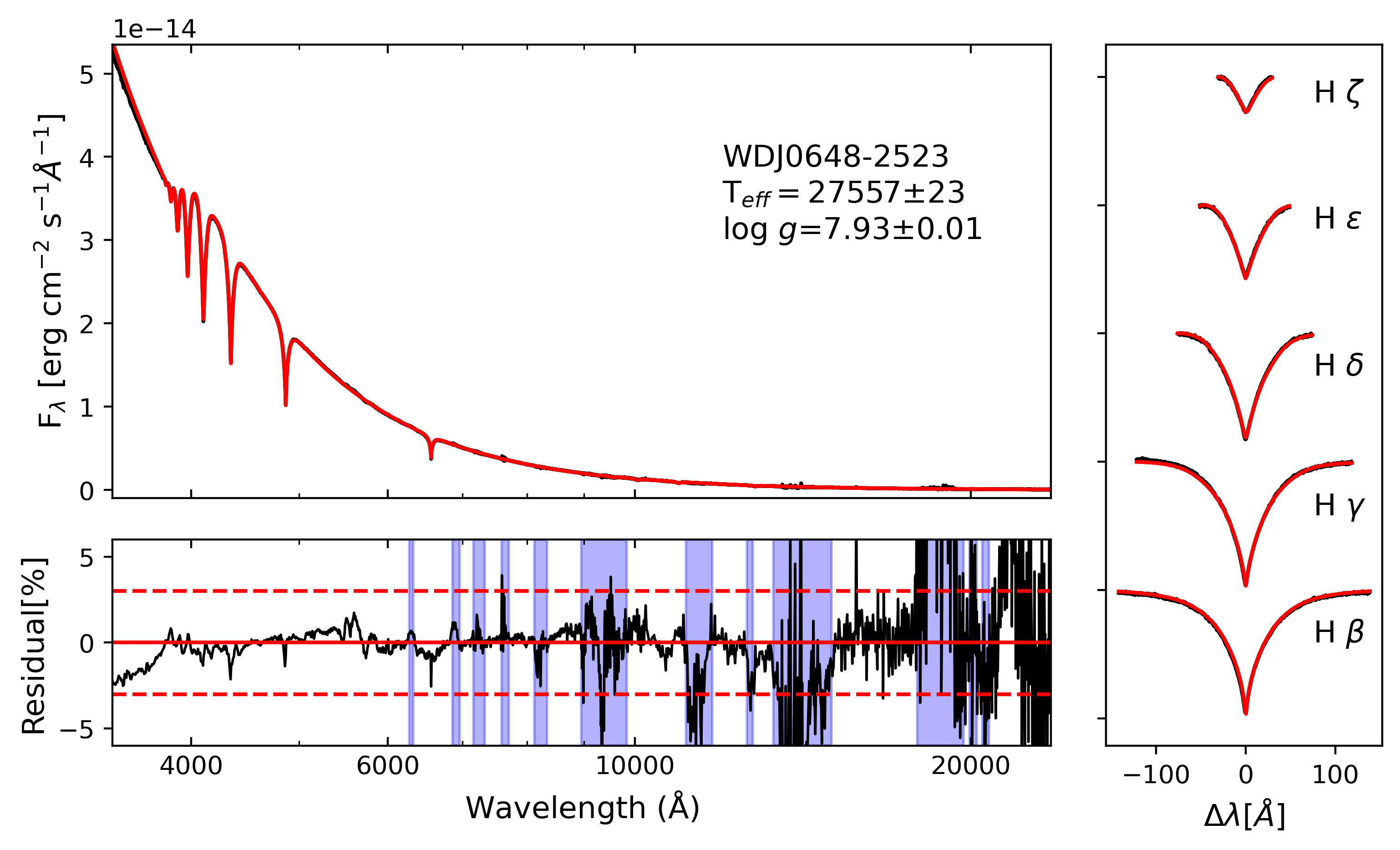}
\caption{Continued.}
\end{center}
\end{figure*}

\setcounter{figure}{0}
\begin{figure*}[th]
\begin{center}
\includegraphics[width=.99\columnwidth]{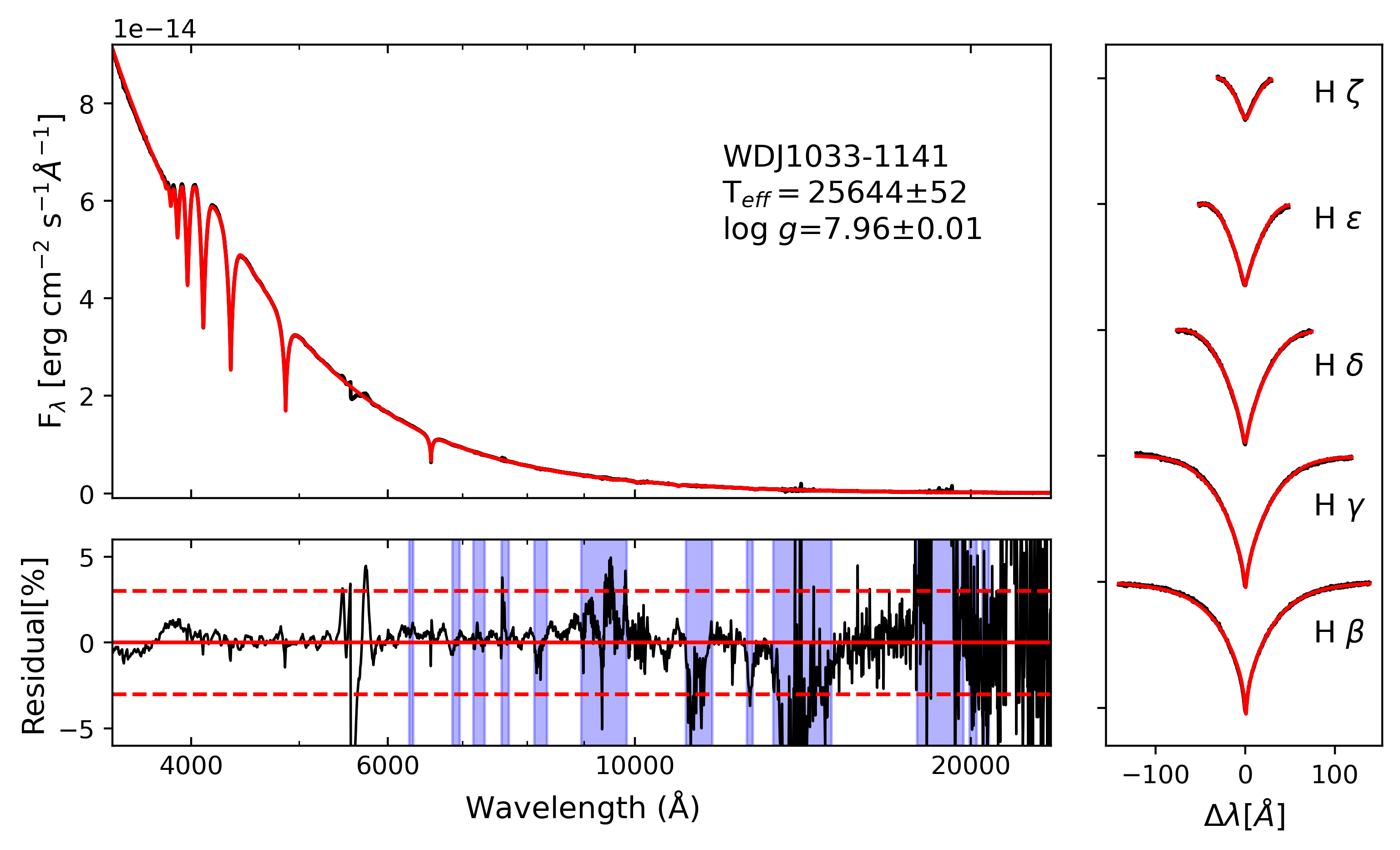}
\includegraphics[width=.99\columnwidth]{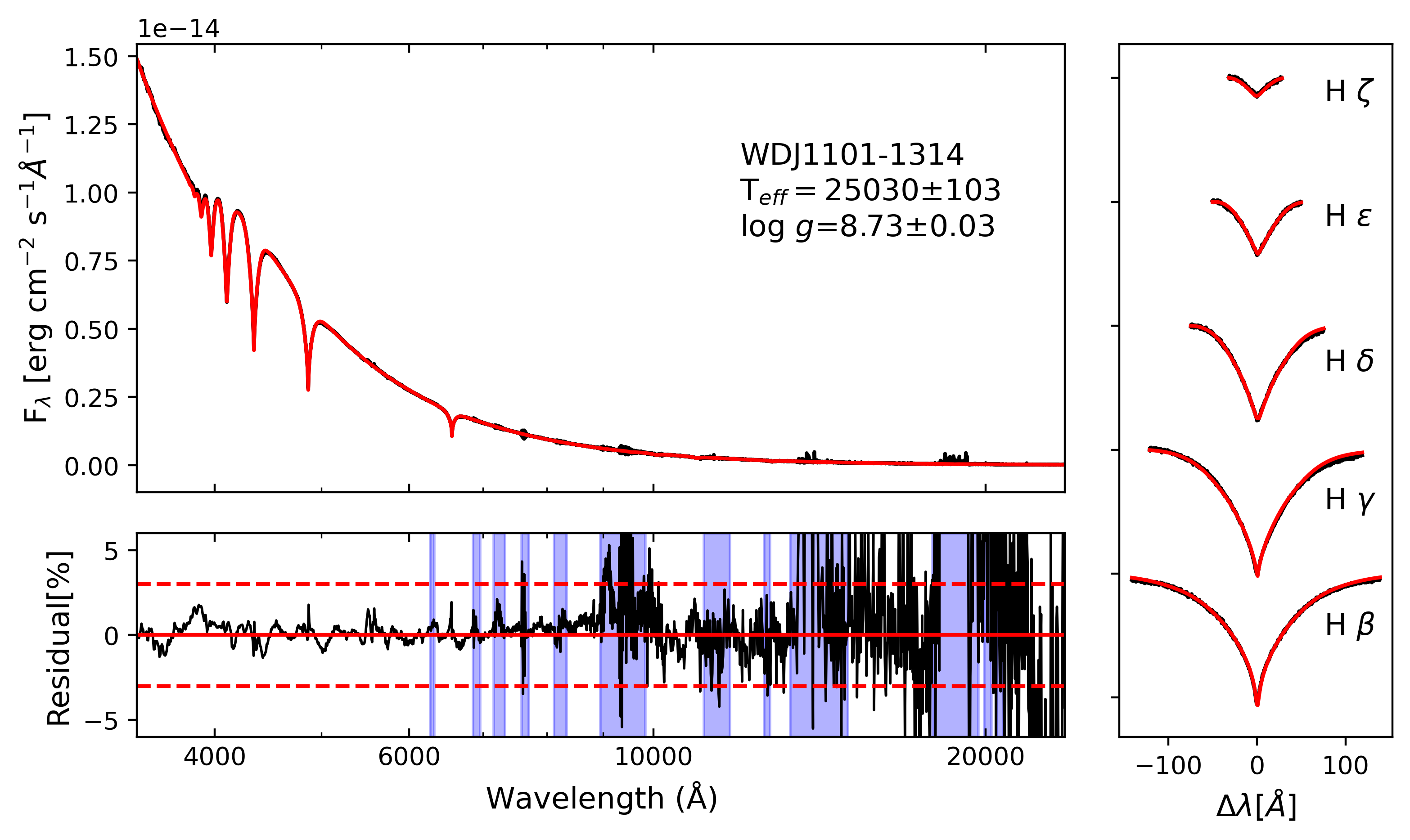}
\caption{Continued.}
\end{center}
\end{figure*}

\setcounter{figure}{0}
\begin{figure*}[th]
\begin{center}

\includegraphics[width=.99\columnwidth]{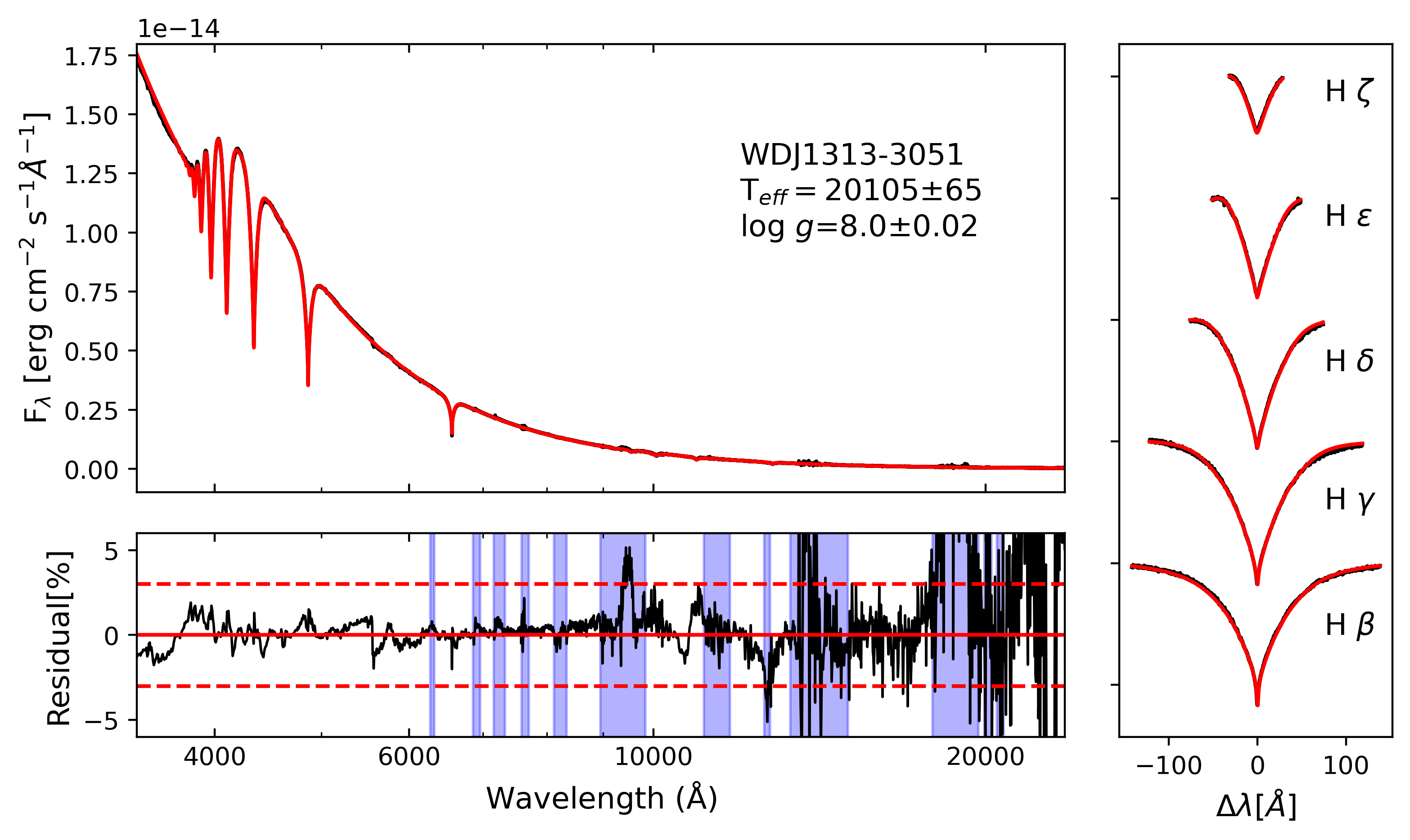}
\includegraphics[width=.99\columnwidth]{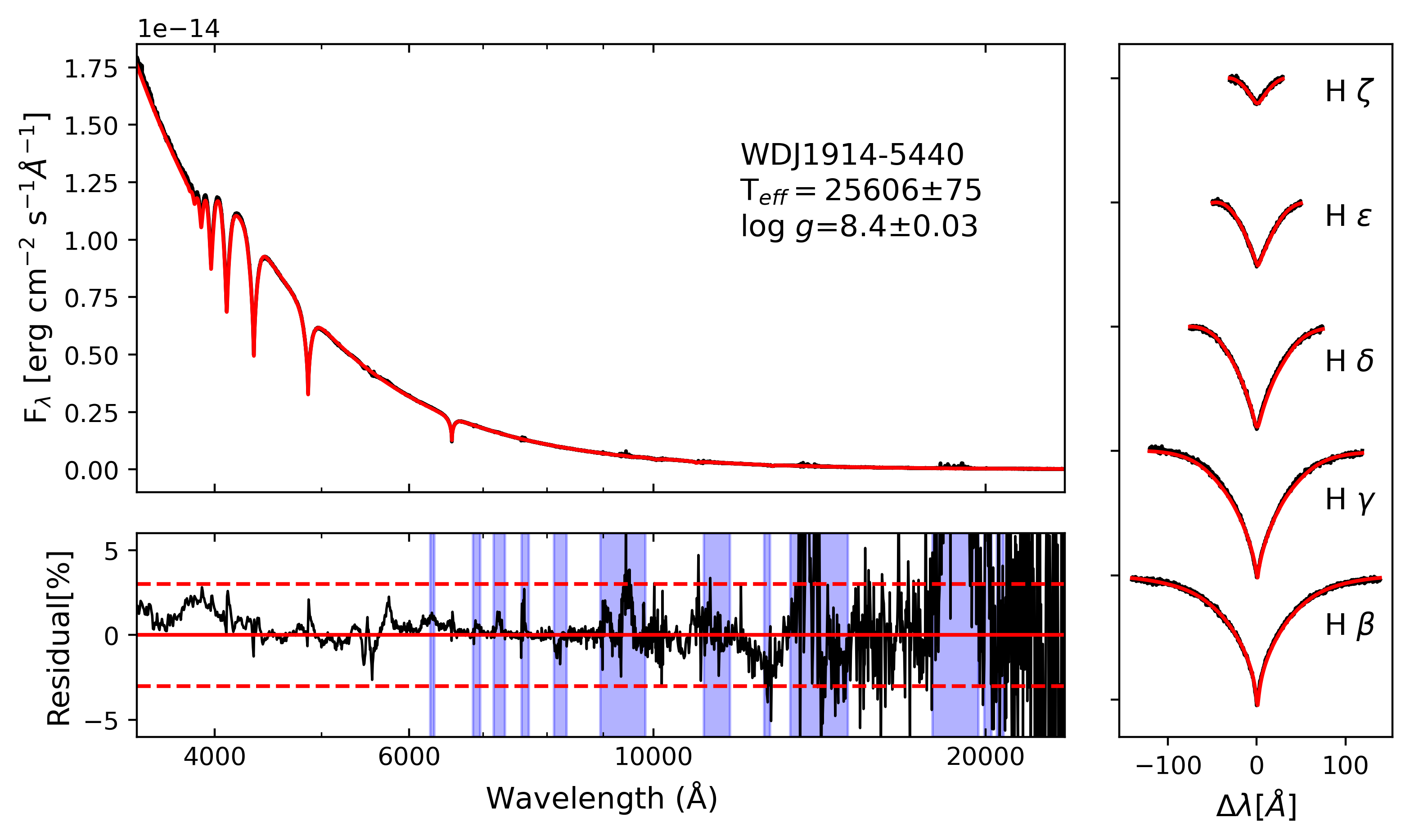}
\caption{Continued.}
\end{center}
\end{figure*}

\setcounter{figure}{0}
\begin{figure*}[th]
\begin{center}

\includegraphics[width=.99\columnwidth]{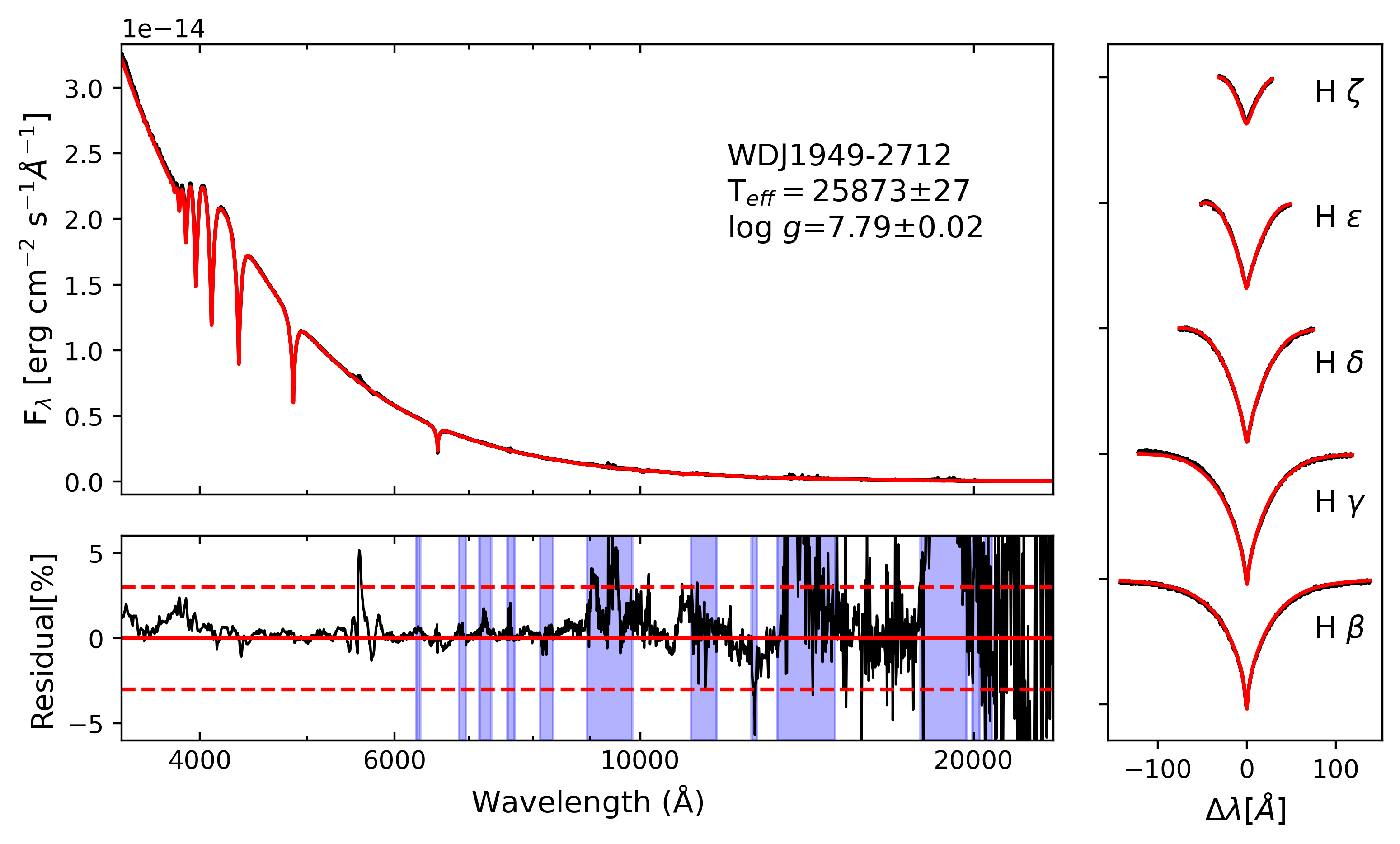}
\includegraphics[width=.99\columnwidth]{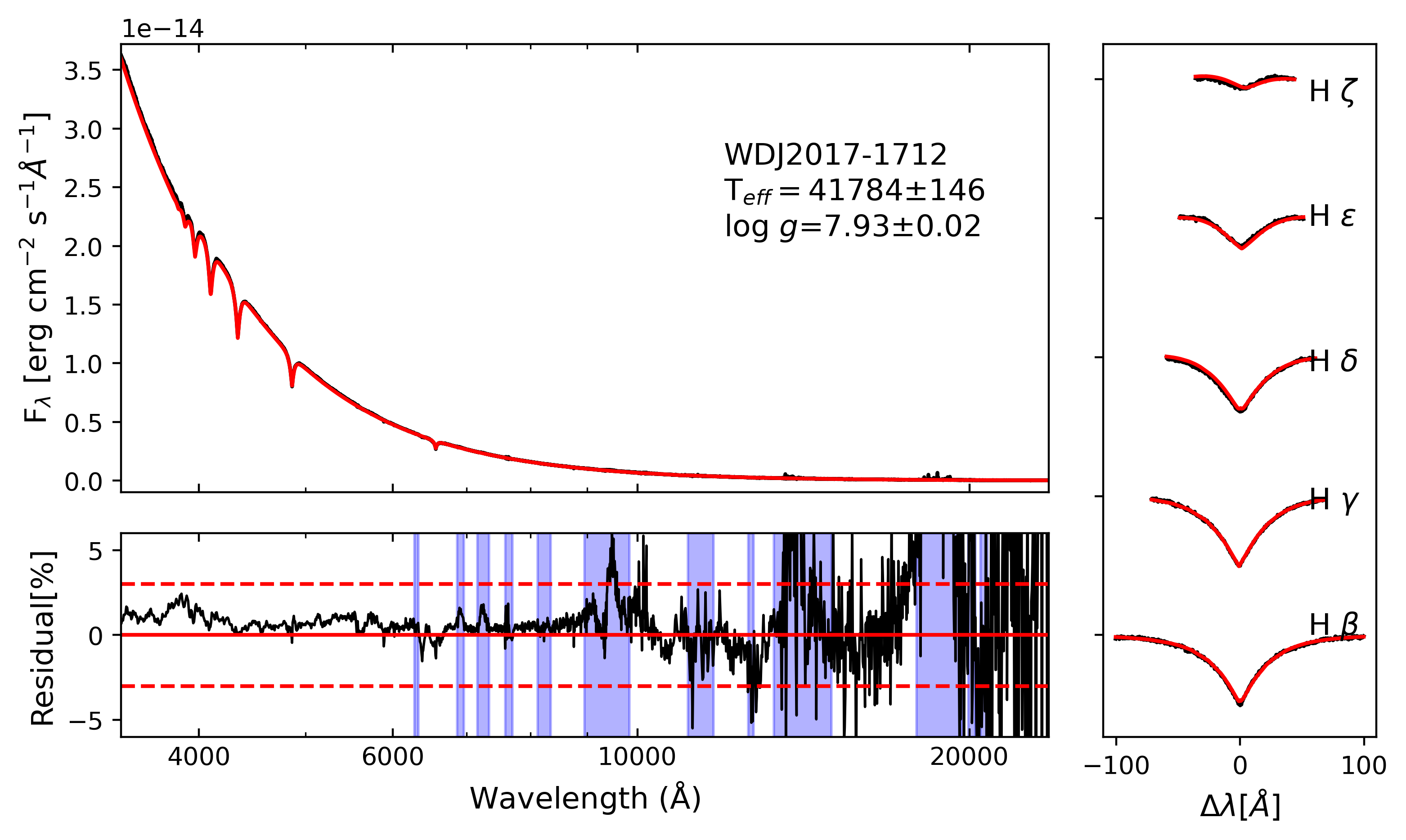}

\caption{Continued.}
\end{center}
\end{figure*}

\clearpage

\setcounter{figure}{0}
\begin{figure*}[th]
\begin{center}
\includegraphics[width=.99\columnwidth]{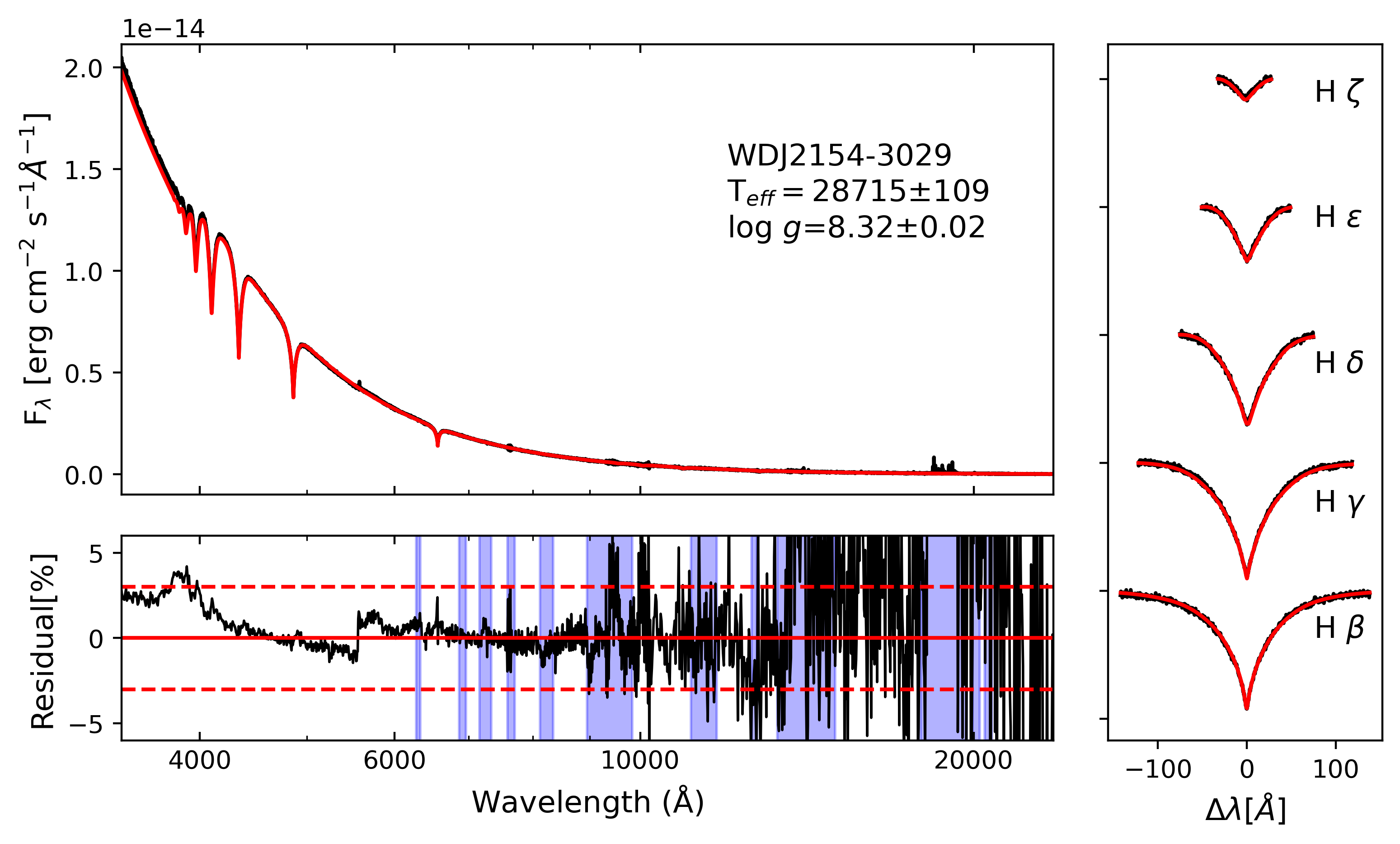}

\caption{Continued.}
\end{center}
\end{figure*}

\section{Candidates rejected as flux standard stars}\label{appendix:B}

\begin{sidewaystable*}[bh]
\centering
\caption[]{Observed targets that were excluded as valid flux standard stars.} \label{tab:rejected}
\setlength{\tabcolsep}{2mm}
\begin{tabular}{lllrrrrp{9.cm}}
\hline\hline
Object & $K$ & $G_{RP}$ & $R.A.$(J2000) & $Dec.$(J2000) & $PM_{R.A.}$  & $PM_{Dec}$ &  Comment\\
\cline{2-3} \cline{6-7}
\textit{Gaia} DR3 ID & \multicolumn{2}{c}{[Vegamag]} & [h m s] & [$^\circ$ ' ''] & \multicolumn{2}{c}{[mas\,yr$^{-1}$]} & \\ 
\hline
WDJ005317.43$-$325956.52 & 14.318 & 13.63 & 00 53 17.43 & $-$32 59 56.52 &  $-$31.138 &   27.509  & colour-dependent residual with model\\
~~~5006486048001153792\\[1mm]
WDJ013158.29$-$683114.51 & 15.32 & 15.04 & 01 31 58.29 & $-$68 31 14.51 &   19.958 &  $-$16.296  & mismatch between photometric and spectroscopic parameters, \\
~~~4691571967756922368 & & & & & & & 
low photometric log $g$ potential double degenerate\\[1mm]
WDJ060020.89$-$101404.50 &   15.893 & 15.01 & 06 00 20.89 & $-$10 14 04.50 &   76.419 &  $-$26.614  & mismatch between photometric and spectroscopic parameters, \\
~~~2999314205559574272 & & & & & & & 
low photometric log $g$, potential double degenerate\\[1mm]
WDJ101952.36$-$140734.26 & 15.277 & 14.81 & 10 19 52.36 & $-$14 07 34.25 &  $-$48.026 &    8.881  & colour-dependent residual; mismatch between photometric and\\
~~~3753200224361795968 & & & & & & & spectroscopic parameters, low photometric log $g$\\[1mm]
WDJ161419.13$-$083326.60 & 14.880 & 14.24 & 16 14 19.13 & $-$08 33 26.61 &   $-$9.285 &  $-$35.908  & mismatch between photometric and spectroscopic parameters,\\
~~~4345767555000692096 & & & & & & & potential double degenerate\\[1mm]
WDJ161755.24$-$153552.15 & 14.318 & 13.68 & 16 17 55.24 & $-$15 35 52.15 & $-$172.259 & $-$164.355  & weak helium lines\\
~~~4328293291578459392\\[1mm]
WDJ170625.36$-$762304.95 & ... & 15.33       & 17 06 25.36 & $-$76 23 04.95 &  $-$40.081 &  $-$36.696     & close cool neighbour\\
~~~5777828673621336832\\[1mm]
WDJ174435.13$-$725935.67 & 12.800 & 15.06 & 17 44 35.13 & $-$72 59 35.66 &   14.480 &   17.398  & spectrum contaminated by close neighbour\\
~~~5803547624984209792\\[1mm]
WDJ195219.68$-$384614.05 & 15.181 & 14.89 & 19 52 19.68 & $-$38 46 14.05 &    5.057 &  $-$58.113  & weak helium lines, possible emission in H$\beta$\\
~~~6690659513515872000\\[1mm]
WDJ195347.76$-$430713.92 & 15.805 & 15.28 & 19 53 47.76 & $-$43 07 13.92 &   109.645 &  $-$80.340 &  mismatch between photometric and spectroscopic parameters\\
~~~6685284241684914688\\
\hline
\end{tabular}
\end{sidewaystable*}

\end{appendix}

\end{document}